\documentclass[11pt,a4paper]{article}
\pdfoutput=1
\usepackage[utf8]{inputenc}
\usepackage[english]{babel}
\usepackage{amsmath, amssymb}
\usepackage{graphicx}
\usepackage{color}
\usepackage{float}
\usepackage{xspace}
\usepackage{cancel}
\usepackage{jcappub}

\newcommand{\vew}{v_{\text{EW}}}
\newcommand{\gyr}{\,\text{Gyrs}}
\newcommand{\s}{\,\text{s}}
\newcommand{\dx}{\,\text{d}}

\newcommand{\lcdm}{\ensuremath{\Lambda}CDM\xspace}

\newcommand{\dcdm}{\ensuremath{_{\mathrm{dcdm}}} }
\newcommand{\wdm}{\ensuremath{_{\mathrm{wdm}}} }
\newcommand{\dr}{\ensuremath{_{\mathrm{dr}}} }

\subheader{\small
	\vspace*{-1.28cm}
	\begin{flushright}
		TUM-HEP-1502/24
	\end{flushright}
}

\title{
Minimal decaying dark matter:\\ from cosmological tensions\\
to neutrino signatures}

\author{Lea Fuß,}
\author{Mathias Garny,}
\author{Alejandro Ibarra}

\affiliation{Technical University of Munich,  
TUM School of Natural Sciences,
Department of Physics,\\
	James-Franck-Str.\ 1, 
	85748 Garching, Germany}

\emailAdd{lea.fuss@tum.de}
\emailAdd{mathias.garny@tum.de}
\emailAdd{alejandro.ibarra@tum.de}

\abstract{%
The invisible decay of cold dark matter into a slightly lighter dark sector particle on cosmological time-scales has been proposed as a solution to the $S_8$ tension. In this work we discuss the possible embedding of this scenario within a particle physics framework, and we investigate its phenomenology. We identify a minimal dark matter decay setup that addresses the $S_8$ tension, while avoiding the stringent constraints from indirect dark matter searches. In our scenario, the dark sector contains two singlet fermions $N_{1,2}$, quasi-degenerate in mass, and carrying lepton number so that the heaviest state ($N_2$) decays into the lightest ($N_1$) and two neutrinos via a higher-dimensional operator  $N_2\to \bar N_1\nu\nu$. The conservation of  lepton number, and the small phase-space available for the decay, forbids the decay channels into hadrons and strongly suppresses the decays into photons or charged leptons. We derive complementary constraints on the model parameters from neutrino detectors, freeze-in dark matter production via $\nu\nu\to N_1N_2$, collider experiments and blazar observations, and we show that the upcoming JUNO neutrino observatory could detect signals of dark matter decay for model parameters addressing the $S_8$ tension if the dark matter mass is below $\simeq 1$\,GeV.
}

\begin{document}

\maketitle
\clearpage

\section{Introduction}
\label{sec:intro}

The \lcdm model describes with remarkable accuracy numerous cosmological observations, including the anistropies in the cosmic microwave background (CMB) and the large-scale structure (LSS) of our Universe. On the other hand, a few observations seem to be in tension with the \lcdm model and may indicate the necessity of an extension. The most conspicuous tension is the discrepancy between the values of the Hubble constant inferred from early and  late Universe observables; this is the well known Hubble tension  (see for example~\cite{Schoneberg:2021qvd}). Furthermore, a tension between the amplitudes of density perturbations inferred from primary CMB or from LSS probes has been reported. This is the  so-called $S_8$ tension, where $S_8 = \sigma_8 \sqrt{\Omega_m/0.3}$ (here, $\sigma_8$ describes the matter fluctuations at scales of $8$\,Mpc$/h$ and $\Omega_m$ is the matter density parameter). 

More specifically, various LSS measurements, including {\it e.g.}~weak lensing shear, galaxy clustering and cluster number counts have reported lower values compared to the one derived from Planck CMB data with $S_8=0.830\pm0.013$~\cite{Planck:2018vyg}. While the significance for each data set is typically only at the level of $1-3\sigma$, they all seem to show a common trend~\cite{Abdalla:2022yfr,DiValentino:2020vvd}.
The strongest deviation was found by the Kilo-Degree Survey KiDS-1000~\cite{KiDS:2020suj} with $S_8=0.759^{+0.024}_{-0.021}$, while the Dark Energy Survey (DES) reports $S_8=0.776^{+0.017}_{-0.017}$~\cite{DES:2021wwk}, and an updated combined analysis of the two obtained $S_8=0.790^{+0.018}_{-0.014}$~\cite{Kilo-DegreeSurvey:2023gfr}. Despite efforts to explain the $S_8$ tension with baryonic or systematic effects it is not easily resolved~\cite{Amon:2022azi,Preston:2023uup}. Another strategy to address the tension is to go beyond \lcdm by changing the dark sector to achieve a suppression of dark matter (DM) clustering, see \emph{e.g.}~\cite{Abdalla:2022yfr}. Even though some models are able to lower the $S_8$ value and thus decrease the tension, there is no definite preference over \lcdm for any of them yet. New surveys, like Euclid~\cite{Euclid:2021icp}, DESI~\cite{DESI:2016fyo} and the LSST survey at the Vera C. Rubin observatory~\cite{LSST:2008ijt}, will probe the amplitude of density perturbations on a wide range of scales and redshifts, and may elucidate in the near feature whether the $S_8$ tension is real.

A promising scenario that addresses the $S_8$ tension is decaying cold dark matter (DCDM). In this scenario, a cold dark matter (CDM) particle decays 
into invisible final states on cosmological time-scales. The cosmological signatures have been studied extensively for massless secondaries~\cite{Audren:2014bca,Enqvist:2015ara,Poulin:2016nat,Enqvist:2019tsa,Nygaard:2020sow,Alvi:2022aam,Berezhiani:2015yta,Bringmann:2018jpr,Pandey:2019plg,DES:2020mpv} as well as for massive ones~\cite{Peter:2010sz,Aoyama:2011ba,Wang:2012eka,Wang:2014ina,Blackadder:2014wpa,Aoyama:2014tga,Blackadder:2015uta,Vattis:2019efj,PhysRevD.103.043014,Haridasu:2020xaa,Abellan:2020pmw,FrancoAbellan:2021sxk,Davari:2022uwd,Simon:2022ftd,Fuss:2022zyt,Holm:2022kkd,Bucko:2023eix}, including probes from the CMB as well as large- and small-scale structure like baryon acoustic oscillations, galaxy clustering, weak lensing, the Lyman-$\alpha$ forest and Milky Way satellites.
Here we focus on a scenario in which the CDM particle is quasi-degenerate in mass with one of the daughter particles in the decay, but with a mass difference that allows the other decay products to be relativistic, being one of the setups that has received increased attention in the context of the $S_8$ tension lately~\cite{Peter:2010sz,Aoyama:2011ba,Wang:2012eka,Wang:2014ina,Blackadder:2014wpa,Aoyama:2014tga,Blackadder:2015uta,Vattis:2019efj,PhysRevD.103.043014,Haridasu:2020xaa,Abellan:2020pmw,FrancoAbellan:2021sxk,Davari:2022uwd,Simon:2022ftd,Fuss:2022zyt,Holm:2022kkd,Bucko:2023eix}. In this scenario, there is a conversion of the rest energy of the mother particle into kinetic energy of the heaviest massive daughter, which gradually builds up a population of warm dark matter (WDM) particles coexisting with the population of CDM particles. 
This leads to a mild suppression of the matter power spectrum on small scales and at late times. The amount and scale of power suppression depend on the lifetime $\tau$ as well as the available fraction of kinetic energy $\epsilon$, respectively. Typical values for alleviating the $S_8$ tension are $\tau \simeq \mathcal{O}(10-100)$\,Gyrs and $\epsilon \simeq 10^{-2}- 10^{-3}$ ~\cite{Abellan:2020pmw,FrancoAbellan:2021sxk,Simon:2022ftd,Fuss:2022zyt,Bucko:2023eix}.

In this work, we identify a minimal embedding of the DCDM scenario within a particle physics framework (see~\cite{Bell:2010fk,Bell:2010qt,Hamaguchi:2017ihw,Bae:2018mgq,Choi:2021uhy,Deshpande:2023gij,Obied:2023clp} for related works). We first discuss the necessary elements of a DCDM scenario which addresses the $S_8$ tension while complying naturally with the stringent limits from gamma-ray and cosmic-ray observations. We then construct a gauge-invariant and Lorentz-invariant operator leading to DCDM, show that it is the simplest one under certain assumptions, and discuss the associated phenomenology in the Early Universe, as well as the possible signatures in laboratory experiments or in astrophysical observations. 

The paper is structured as follows: In Sec.~\ref{sec:Overview DCDM}, we give a brief overview of DCDM and its existing phenomenology and cosmological signatures. Then, the concrete minimal model is developed and motivated in Sec.~\ref{sec:minimalmodel}, including an evaluation of the main decay channel. Next, in Sec.~\ref{sec:NeutrinoConstraints} we derive limits from diffuse neutrino flux measurements. In Sec.~\ref{sec:Production} we consider freeze-in production of DM within the minimal model. In Sec.~\ref{sec:other_signatures} we discuss complementary signatures, including higher-order decays into charged particles and gamma-rays, invisible Higgs decay, as well as neutrino flux attenuation from blazars due to neutrino-DM scatterings. We conclude in Sec.~\ref{sec:Conclusion}.
The appendices provide further technical material, relevant for the computation of decay rates, for freeze-in, neutrino-DM scattering, as well as a discussion of the limit of very low DM masses.

\section{The decaying cold dark matter scenario and the \texorpdfstring{$\mathbf{S_8}$}{S8} tension}
\label{sec:Overview DCDM}

The simplest DCDM scenario addressing the $S_8$ tension consists of a population of initially cold dark matter (DCDM) decaying into one massive and one massless particle species, that act as warm dark matter (WDM) and dark radiation (DR), respectively~\cite{Abellan:2020pmw,FrancoAbellan:2021sxk,Simon:2022ftd,Fuss:2022zyt,Bucko:2023eix}
\begin{equation}
	\text{DCDM} \ \rightarrow \ \text{WDM} \ + \ \text{DR}\,.
\end{equation}
From the point of view of cosmology all ``dark'' particle species are assumed to have negligible interactions with visible matter at the relevant time-scales, \emph{i.e.}~during and after the recombination epoch. We note that the Standard Model (SM) neutrinos satisfy all the requirements to be ``dark radiation'' with this definition, as we will emphasize below.

The effects of DCDM on cosmology can be entirely captured by two parameters: the decay rate $\Gamma$ (or equivalently the lifetime $\tau=\Gamma^{-1}$), and the relative mass splitting $\epsilon$, defined as 
\begin{equation}\label{eq:splitting}
    \epsilon \equiv \frac{1}{2}\left(1 - \frac{m^2}{M^2} \right)\,,
\end{equation}
where $M$ and $m$ are the masses of the DCDM and the WDM particles, which  determines the fraction of rest mass that is converted into kinetic energy in the decay.
In this paper, we focus on the limit when the DCDM and the WDM particles are quasi-degenerate in mass, namely $\epsilon\ll 1$, so that there is a gradual ``heating'' of the DM as more and more CDM particles decay, and that is investigated in view of the $S_8$ tension~\cite{Peter:2010sz,Aoyama:2011ba,Wang:2012eka,Wang:2014ina,Blackadder:2014wpa,Aoyama:2014tga,Blackadder:2015uta,Vattis:2019efj,PhysRevD.103.043014,Haridasu:2020xaa,Abellan:2020pmw,FrancoAbellan:2021sxk,Davari:2022uwd,Simon:2022ftd,Fuss:2022zyt,Holm:2022kkd,Bucko:2023eix}.

The Boltzmann equations for this setup include source and loss terms for the respective new dark species. At the homogeneous and isotropic background level, they are given by~\cite{Abellan:2020pmw}
\begin{flalign}
	\dot{\bar{f}}\dcdm(q,\tau) &= -a\Gamma\bar{f}\dcdm(q,\tau) \,,\nonumber  \\
	\dot{\bar{f}}\wdm(q,\tau) = \dot{\bar{f}}\dr(q,\tau) &= \frac{a\Gamma\bar{N}\dcdm}{4\pi q^2} \delta(q- ap_\text{2-body})\,,
	\label{eq:BE}
\end{flalign}
where $f$ are the respective phase-space distribution functions, $q=ap$ is the co-moving momentum, $a$ the  scale-factor and $\bar{N}\dcdm$ the number density of the mother particle, which drops exponentially in time. A dot denotes derivative with respect to conformal time $\eta$, related to the usual cosmic time $t$ via $d\eta=dt/a$.
Multiplying the Boltzmann equations with the respective particle energies and integrating over all momentum modes yields equations for the average energy densities denoted by $\bar{\rho}$,
\begin{flalign}
	\dot{\bar{\rho}}\dcdm &= -3\mathcal{H} \bar{\rho}\dcdm - a\Gamma\bar{\rho}\dcdm\,, \nonumber\\
	\dot{\bar{\rho}}\dr &= -4\mathcal{H} \bar{\rho}\dr + \epsilon a\Gamma\bar{\rho}\dcdm\,, \nonumber\\
	\dot{\bar{\rho}}\wdm &= -3(1+\omega)\mathcal{H} \bar{\rho}\wdm + (1-\epsilon) a\Gamma\bar{\rho}\dcdm\,,
	\label{eq:densityDE}
\end{flalign}
with ${\mathcal H}=aH$ where $H$ is the Hubble rate, and with the equation-of-state parameter $\omega=\bar{P}\wdm/\bar{\rho}\wdm$ for WDM where $\bar{P}\wdm$ is the average pressure. On top of the usual cosmological evolution, the terms involving $\Gamma$ describe the impact of the decay on the background densities.

The phenomenology of this scenario has been discussed in various works, \emph{e.g.}~\cite{Bucko:2023eix,Simon:2022ftd,Fuss:2022zyt,DES:2022doi,Abellan:2020pmw,Peter:2010sz}. The most prominent implication is a suppression of the matter power spectrum on small scales, being linked to the $S_8$ tension. Since the massive decay product receives a velocity kick, part of the DM develops a non-zero pressure and acts like a warm DM component building up over time. Compared to CDM, the additional WDM can more easily escape overdensities and wash out structure on small physical scales. This leads to a gradual suppression of the power spectrum for large wavenumbers $k$. Since $\epsilon$ determines the momentum of the WDM, it sets the free-streaming scale and thus the wavenumber at which the suppression starts. In contrast, the decay time $\tau$ controls the fraction of WDM at any given redshift $z$ and is responsible for the amount of suppression. Notably, the decay implies a pronounced redshift-dependence of the amplitude of suppression.

Another related effect of the velocity kick is the difference in halo evolution compared to CDM. For small enough halos, with typical virial velocities of the order of or below the kick velocity,  WDM particles are able to disrupt or escape the halo. This suppresses the number of small halos compared to the $\Lambda$CDM model~\cite{Peter:2010sz,DES:2022doi}. Apart from matter fluctuations, DCDM can change the cosmological background evolution as described in Eq.\,\eqref{eq:BE}. However, this effect is negligible for sufficiently large $\tau$ or small $\epsilon$. For $10^5\,\text{yr}\ll \tau \ll t_0$ and sizeable $\epsilon$, the extra DR increases the distance to the sound horizon which then requires a higher $\Omega_\Lambda$ (and thus $H_0$) to not shift the acoustic peaks in the CMB anisotropy spectrum. While this possibility has been discussed in context of the Hubble tension, it is highly constrained by supernova, baryonic acoustic oscillation (BAO) and CMB data~\cite{Nygaard:2023gel,Simon:2022ftd}.

\begin{figure}[t]
    \centering
    \includegraphics[width=0.6 \textwidth ]{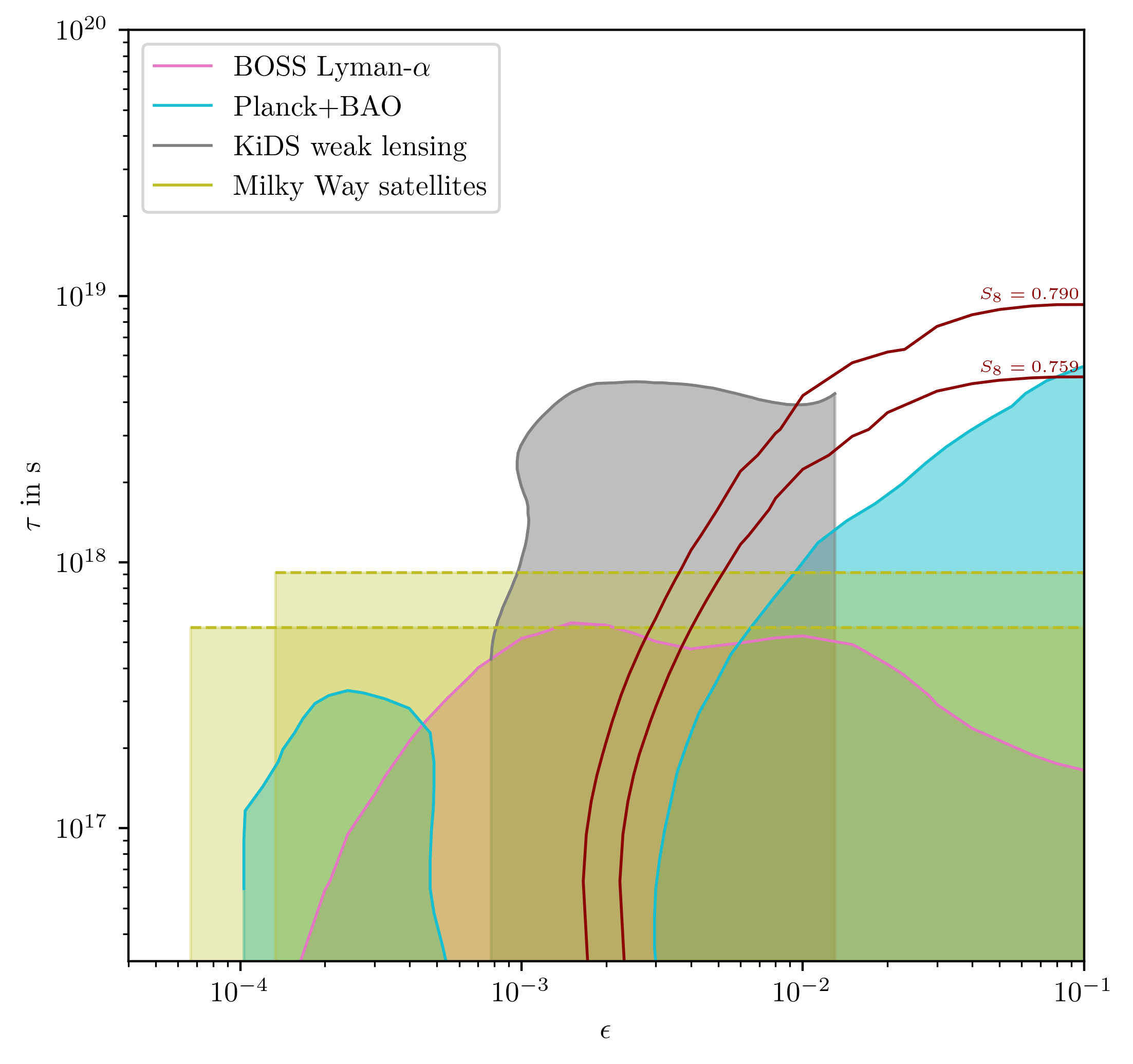}
    \caption{Allowed parameter space of the decaying cold dark matter scenario $\text{DCDM} \ \rightarrow \ \text{WDM} \ + \ \text{DR}$. The region between the two thick red lines 
    highlights the region in parameter space addressing the $S_8$ tension, and the shaded regions show various constraints derived from Lyman-$\alpha$, CMB, weak lensing and Milky Way satellite data (see legend and main text for details).
    }
    \label{fig:cosmo_constraints}
\end{figure}

Overall, DCDM with $\tau\gtrsim t_0$ and $\epsilon\ll 1$ can primarily be probed by LSS data as well as halo abundances and properties. In Fig.~\ref{fig:cosmo_constraints}, we show a collection of cosmological constraints on DCDM in the $\epsilon-\tau$ plane, where the upper left corner converges to \lcdm. In blue, limits from combining Planck CMB, BAO and Pantheon supernova data as obtained in~\cite{Simon:2022ftd} are shown.
Note that this result was actually reported as a confidence interval around the best fit value, that we inverted to indicate which values are approximately excluded. In pink, we show limits from~\cite{Fuss:2022zyt}, where the Lyman-$\alpha$ flux power spectrum measured by BOSS at $z=3.0-4.2$~\cite{Chabanier:2018rga} was used to constrain the matter power spectrum.
In gray, we display results from a weak lensing shear analysis performed in~\cite{Bucko:2023eix}, leading to rather strong constraints. In this work, an emulator was trained to determine the power spectrum of DCDM on non-linear scales, including baryonic effects, in order to analyze KiDS-1000~\cite{KiDS:2020suj} combined with Planck CMB data~\cite{Planck:2018vyg}.
Finally, in olive we show constraints from~\cite{Peter:2010sz} derived from Milky Way satellite abundances and the respective halo mass functions for two different velocity kicks that are then extrapolated to higher values.

For illustration, the dark red contour lines in Fig.~\ref{fig:cosmo_constraints} show the $(\epsilon,\tau)$ parameters for which DCDM can reproduce the $S_8$ values reported by KiDS, $S_8=0.759$~\cite{KiDS:2020suj}, and KiDS+DES, $S_8=0.79$~\cite{Kilo-DegreeSurvey:2023gfr}, respectively. All other cosmological model parameters are set to the Planck 2018 best-fit values~\cite{Planck:2018vyg} within $\Lambda$CDM to obtain the contour lines, motivated by the fact that DCDM and $\Lambda$CDM are indistinguishable at times $t\ll\tau$, \emph{i.e.}~in particular around recombination. For comparison, within $\Lambda$CDM $S_8=0.83$ for Planck parameters.
As can be seen in Fig.~\ref{fig:cosmo_constraints}, current cosmological constraints on DCDM still leave an open window where the $S_8$ value can be reduced. Typical values in this window are of the order of $\epsilon\simeq 0.01$ and $\tau \simeq 100 \gyr \simeq 3\cdot 10^{18}\s$.

This simple setup can be extended to include more DR particles in the final state, for instance the three-body DCDM decay~\cite{Fuss:2022zyt,Blackadder:2014wpa}
	\begin{equation}
		\text{DCDM} \ \rightarrow \ \text{WDM} \ + \ \text{DR}_a\ + \ \text{DR}_b\,.
	\end{equation} 
As for the two-body decays, the cosmological implications of the decay can be encoded in the decay rate,  $\Gamma$, and in the fraction of the energy of the mother particle that is converted into kinetic energy, determined by the mass splitting $\epsilon$ in Eq.\,\eqref{eq:splitting}.
On the other hand, this scenario presents the complication that the decay products have a continuous momentum distribution, $d\Gamma/d q_i$, with $q_i$ the momenta of the different daughter particles, and which is highly model dependent. The Boltzmann equations at the background level then read~\cite{Fuss:2022zyt}
	\begin{flalign}
		\dot{\bar{f}}\dcdm(q,\tau) &= -a\Gamma\bar{f}\dcdm(q,\tau)\,, \nonumber \\
		\dot{\bar{f}}\wdm(q_1,\tau)  &= \frac{a\Gamma\bar{N}\dcdm}{4\pi q_1^2} \left(\frac{1}{\Gamma}\frac{d\Gamma}{dq_1}\right)\,,\nonumber\\
		\dot{\bar{f}}_{\text{dr}_{a,b}}(q_{2,3},\tau) &=\frac{a\Gamma\bar{N}\dcdm}{4\pi q_{2,3}^2} \left(\frac{1}{\Gamma}\frac{d\Gamma}{dq_{2,3}}\right)\,.
		\label{eq:BE3body}
	\end{flalign}
Compared to Eq.\,\eqref{eq:BE} the Dirac delta is replaced by the momentum distribution.
Multiplying these equations by energy and integrating over momentum yields evolution equations for the energy density, thus generalizing Eq.\,\eqref{eq:densityDE} to three-body decays
\begin{flalign}
		\dot{\bar{\rho}}\dcdm &= -3\mathcal{H} \bar{\rho}\dcdm - a\Gamma\bar{\rho}\dcdm\,, \nonumber\\
		\dot{\bar{\rho}}\dr &= -4\mathcal{H} \bar{\rho}\dr + (2\langle E_\text{dr}\rangle/M) a\Gamma\bar{\rho}\dcdm\,, \nonumber\\
		\dot{\bar{\rho}}\wdm &= -3(1+\omega)\mathcal{H} \bar{\rho}\wdm + (\langle E_\text{wdm}\rangle/M) a\Gamma\bar{\rho}\dcdm\,,
		\label{eq:densityDE3body}
	\end{flalign}
where  the average over the momentum distribution is denoted by $\langle X \rangle = 1/\Gamma \cdot \int X \dx\Gamma$.
Here $\bar{\rho}\dr$ is the sum of the energy densities of the two DR contributions, which evolve in the same way.

In~\cite{Fuss:2022zyt} it was argued that in the limit $\epsilon\ll 1$ primarily interesting to us, the impact of the three-body decay on cosmological observables can be effectively mapped on an equivalent two-body decay model.
We review the reasoning in the following. The impact of dark matter decay on the background evolution becomes negligible for $\epsilon\to 0$ since then $\left\langle 2 E\dr\right\rangle/M = \mathcal{O}(\epsilon)$ and $\left\langle E\wdm\right\rangle/M = 1 + \mathcal{O}(\epsilon)$. This implies that $\bar{\rho}\dr$ becomes negligibly small and the total dark matter density $\bar{\rho}\dcdm+\bar{\rho}\wdm$ evolves approximately as if there was no decay. The same argument applies to the two-body decay, such that for both cases the background evolution is unaltered compared to \lcdm for $\epsilon\ll 1$. Instead, the perturbations are responsible for the dominant effect on cosmological observables, capturing the heating of the WDM component produced in the decay. A fluid approximation for the WDM component that keeps track of the first two moments of the distribution function was introduced and validated against the full Boltzmann hierarchy for $\epsilon\ll 1$ in~\cite{Abellan:2020pmw}. The impact of the decay
is in this framework dominantly captured by an effective sound velocity, related to the adiabatic value $c_g^2\equiv \dot{\bar P}\wdm/\dot{\bar\rho}\wdm$. For the three-body decay, it can be expressed as~\cite{Fuss:2022zyt}
\begin{equation}\label{eq:soundvelocity}
		\begin{aligned}
			c_g^2 =  &\left(\omega\left(5- \frac{\mathfrak{p}}{\bar{P}\wdm} \right) - a\Gamma \frac{\bar{\rho}\dcdm}{\bar{\rho}\wdm} \frac{1}{\mathcal{H} M} \left\langle \frac{p\wdm^2}{3E\wdm} \right\rangle \right) \\
			\cdot &\left( 3(1+\omega) - a\Gamma \frac{\bar{\rho}\dcdm}{\bar{\rho}\wdm} \frac{1}{\mathcal{H} M} \left\langle E\wdm\right\rangle \right)^{-1}\,,
		\end{aligned}
	\end{equation}
where $\mathfrak{p}$ is the pseudo-pressure~\cite{Lesgourgues:2011rh}. 
The main difference to two-body decays is the appearance of averages involving the energy $E\wdm$ and momentum $p\wdm$ of the massive decay product over the decay spectrum, capturing the model-dependence. The two-body case is recovered by replacing $\langle E\wdm\rangle \to E_\text{2-body} = (1-\epsilon)M$
and $\langle p^2\wdm/(3E\wdm)\rangle \to p_\text{2-body}^2/(3E_\text{2-body}) = \epsilon^2 M/(3-3\epsilon)$.

The main idea of the mapping of cosmological constraints from two- to three-body decays is that models with a given value of $c_g^2$ will lead to (approximately) identical predictions of observables such as the matter power spectrum, since their impact is mostly captured by the sound velocity for small $\epsilon$~\cite{Abellan:2020pmw}. We thus consider a fictitious two-body decay model (with fictitious mass splitting $\epsilon'$) that leads to the same sound velocity as the three-body decay model (with actual mass splitting $\epsilon$) of interest. To obtain this mapping, we note that
both $\mathfrak{p}$ and $\omega$ are suppressed with $\mathcal{O}(\epsilon^2)$ so the numerator in Eq.\,\eqref{eq:soundvelocity} is dominated by the second term. Additionally, the denominator is the same for the two- and three-body case at leading order in $\epsilon$ and thus the ratio of the sound velocities for each case can be written as
\begin{equation}
		\frac{c_g^2|_\text{3-body}}{c_g^2|_\text{2-body}} = \frac{\langle p_\text{wdm}^2/3E_\text{wdm}\rangle}{p_\text{2-body}^2/3E_\text{2-body}}\,.
	\end{equation}
For the two-body decay, $p_\text{2-body}^2/3E_\text{2-body} = {\epsilon'}^2M/(3-3\epsilon')\simeq {\epsilon'}^2 M/3$, while a more complicated dependence on $\epsilon$ occurs for the three-body decay, depending on the momentum distribution. Mapping cosmological constraints from two- to three-body decays corresponds to finding the value of $\epsilon'$ such that
\begin{equation}\label{eq:new_epsilon}
  c_g^2|_\text{3-body}(\epsilon) \stackrel{!}{=} c_g^2|_\text{2-body}(\epsilon')\,.
\end{equation}
Inserting the explicit expressions for the sound velocities yields the desired mapping $\epsilon'(\epsilon)$.
For example, when assuming a constant (momentum-independent) matrix element for the three-body decay, one obtains~\cite{Fuss:2022zyt}
\begin{equation}
	\epsilon'(\epsilon) = \sqrt{\frac{3}{5}}\epsilon + \mathcal{O}(\epsilon^2)\;.
   \label{eq:mapping}
\end{equation}
We will apply in the next section this mapping procedure to a concrete particle physics model.

\section{A minimal model of decaying cold dark matter}
\label{sec:minimalmodel}

The SM does not contain candidates for CDM nor for WDM. Therefore, the model requires at least two new particle species, that have to be pseudo-degenerate in mass. It is plausible to consider that two particles with a small mass difference carry the same spin, therefore we will consider these two new particles to be  Dirac fermions, denoted by $N_{1}$ and $N_2$, with masses of $m_{N_2}=M$ and $m_{N_1}=m \approx M (1-\epsilon)$.  On the other hand, as discussed in Sec.
\ref{sec:Overview DCDM}, the ``dark radiation'' consists of relativistic particles that have negligible interactions with visible matter after the onset of the recombination epoch. Notably, the SM contains particles fulfilling these properties: the active neutrinos. Therefore, in a minimal setup, one can identify the DR particles with the neutrinos. 

To describe the interaction, we will use an effective field theory approach. The lowest dimensional operators involving $N_1$, $N_2$ and SM neutrinos are of the form
\begin{align}
    \mathcal{L}&\sim (\bar{L}N_1)(\bar{N}_2 L)+\text{h.c.}\quad \text{or} \quad
    \mathcal{L}\sim (\bar{L}N_1)(\bar{N}^c_2 L)+\text{h.c.}\,,
    \label{eq:naiveoperators}
\end{align}
where $L=(\nu_L,e_L)$ is the SM lepton doublet, which includes the left-handed neutrino and electron fields. These six-dimensional operators lead to the decays $N_2 \rightarrow N_1 \nu \bar\nu$ and $N_2 \rightarrow \bar N_1 \nu \bar\nu$, and as discussed in  Sec.~\ref{sec:Overview DCDM} can potentially solve the $S_8$ tension if  $\Gamma\simeq 10^{-18}-10^{-19}\s^{-1}$ and  $\epsilon \simeq 10^{-2}- 10^{-3}$.
However, the same operators  also generate decays into charged leptons, $N_2 \rightarrow N_1 e^- e^+$ or $N_2 \rightarrow \bar N_1 e^- e^+$, with comparable rate if kinematically accessible, and also the decay $N_2\rightarrow N_1 \gamma$ at the one-loop level,  with a rate suppressed by a factor ${\cal O}(10^{-3})$~\cite{Garny:2010eg}. Gamma-ray observations restrict the dark matter decay width to be $\Gamma_\gamma\lesssim10^{-30}\s^{-1}$~\cite{Berteaud:2022tws,Calore:2022pks,Essig:2013goa} and positron flux measurements to be $\Gamma_{e^+}\lesssim10^{-28}\s^{-1}$~\cite{DelaTorreLuque:2023olp,DelaTorreLuque:2023cef,Jin:2013nta}. Therefore, solving the $S_8$ tension with these two operators seems at odds with the gamma-ray and the positron observations. 

In order to forbid these operators, we assign lepton number to $N_{1,2}$, corresponding to a global $U(1)$ transformation for which $N_{1}\to e^{ i\alpha}N_{1}$ and $N_{2}\to e^{i\alpha}N_{2}$. In addition, we need to introduce a second global $U(1)$ symmetry, that transforms  $N_{1}\to e^{ i\alpha}N_{1}$ and $N_{2}\to e^{-i\alpha}N_{2}$, while all SM particles transform trivially under this symmetry. We will refer to the charge under this symmetry as  ``N-number''. Note that the conservation of lepton number and the conservation of  ``N-number'', ensure that not only both of the four-fermion interactions from Eq.\,\eqref{eq:naiveoperators} are absent, but also similar four-fermion interactions where the lepton doublets are replaced by any other SM fermion fields. These symmetries are also compatible with Dirac mass terms ${\cal L}_{\text{mass}}=-m_{N_1}\bar N_1N_1-m_{N_2}\bar N_2 N_2$ while forbidding any mass mixing terms (\emph{e.g.}~$\bar N_1 N_2$) or Majorana mass terms (\emph{i.e.}~$\bar N_i^c N_j$), such that $N_1$ and $N_2$ indeed correspond to Dirac fermion mass eigenstates. Finally, the symmetry $N_{1}\to e^{ i\alpha}N_{1}$ ensures the stability of $N_1$, since it is the lightest particle carrying ``N-number''.

The conservation of ``N-number'' and lepton number forbids the dimension-six operators in Eq.\,\eqref{eq:naiveoperators}. However, there exist higher dimensional operators allowed by the symmetries. The simplest one is the dimension-eight operator\footnote{An analogous operator is given by $\bar N_1  N_2^c \bar L^c\tilde H^*\,\tilde H^\dag L$. We focus on Eq.\,\eqref{eq:Lagrangian} for definiteness here. There exist other lower-dimensional operators, \emph{e.g.}~$\bar N_i N_i H^\dag H$, that however are irrelevant for DM decay.}
\begin{equation}
	{\cal L}_{\text{int}} = \frac{1}{\Lambda^4} \left(\bar{L}\tilde{H}P_R N_2\right) \left(\bar{L}\tilde{H}P_R N_1\right) + \text{h.c.}\,,
    \label{eq:Lagrangian}
\end{equation}
which leads after the electroweak symmetry breaking to the four-fermion interaction of $N_{1,2}$ and a neutrino pair, described by the effective Lagrangian
\begin{equation}\label{eq:nunuNN}
	\mathcal{L}_{\text{eff}} = \frac{\vew^2}{2\Lambda^4} \, \bar{\nu} P_R N_2 \, \bar{\nu} P_R N_1 + \text{h.c.}\,.
\end{equation}
This operator induces the decay
\begin{equation}\label{eq:nunudecay}
  N_2\rightarrow \bar N_1 \nu\nu\,,
\end{equation}
involving two neutrinos in the final state, instead of a neutrino-antineutrino pair. We note that the hypothetical decay $N_2\rightarrow \bar N_1 e^-e^-$ is allowed by the conservation of the ``N-number'' and the lepton number, but not by the conservation of the local $U(1)$ electromagnetic symmetry. Further, the decay $N_2\rightarrow \bar N_1 \gamma$ is not allowed by the conservation of the lepton number. Therefore, this scenario is a potentially viable DCDM scenario, since the rate for $N_2\rightarrow N_1 \nu\nu$ could be in the ballpark of the values required to address the $S_8$ tension, while avoiding the constraints from gamma-ray and positron observations. Other, more suppressed, decay channels producing gamma-rays and positrons will be discussed in Sec.\,\ref{sec:other_signatures}.

In order to determine the regions of the parameter space that  is relevant for the $S_8$ tension, we first calculate the squared matrix element for $N_2\rightarrow \bar N_1 \nu\nu$. We find
\begin{equation}\label{eq:matrixelementsquared}
   \overline{|{\cal M}|^2} = \frac{\vew^4}{2\Lambda^8}
	\left( \right. 2(k_1\cdot p_1)(k_2\cdot p_2) +  2(k_2\cdot p_1)(k_1\cdot p_2) -  (k_1\cdot k_2)(p_1\cdot p_2) \left. \right)\,,
\end{equation}
where $k_{1,2}$ are the momenta of the neutrinos, $p_{1,2}$ the momenta of $N_{2,1}$, and we summed over the final state spins and averaged over the initial state spins. The differential decay rate reads, keeping the leading order in an expansion in small $\epsilon$,
\begin{equation}
    \frac{\dx\Gamma_{N_2\rightarrow N_1\nu\nu}}{\dx p\wdm} = \frac{\vew^4}{1536\pi^3 \Lambda^8} p\wdm^2 \left( p\wdm^2+ 3M^2\epsilon^2 \right)\,,
\end{equation}
and the total decay rate, 
\begin{equation}
    \Gamma_{N_2\rightarrow N_1\nu\nu} = \frac{\vew^4}{1280 \pi^3 \Lambda^8}\left( \epsilon M \right)^5\,.
    \label{eq:maindecaywidth}
\end{equation} 
Note the suppression by $\epsilon^5$, arising partially from the marix element ($\epsilon^2$) and the phase-space ($\epsilon^3$).
Thus, the suppression scale $\Lambda$ of the effective interaction to produce a given lifetime $\tau$ is
\begin{equation}
    \label{eq:Lambda}
    \Lambda = \left( \frac{\vew^4}{1280 \pi^3}\tau\left( \epsilon M \right)^5 \right) ^{1/8}
    \approx 12\,\text{TeV}\left(\frac{\tau}{100\text{Gyrs}}\right)^{1/8}\left(\frac{\epsilon M}{\text{MeV}}\right)^{5/8}\,.
\end{equation}

We can finally translate cosmological constraints obtained for a two-body decay using the mapping derived in~\cite{Fuss:2022zyt} and reviewed in Sec.~\ref{sec:Overview DCDM}. For this purpose, we have to calculate
\begin{equation}
		\langle p\wdm^2/3E\wdm\rangle\ = \frac{1}{\Gamma_{N_2\rightarrow N_1\nu\nu}}\int \frac{p\wdm^2}{2 E\wdm} \frac{\dx\Gamma_{N_2\rightarrow N_1\nu\nu}}{\dx p\wdm}\dx p\wdm\,,
	\end{equation}
where $E\wdm=\sqrt{M^2(1-2\epsilon)+p\wdm^2}\simeq M$.
Using Eq.\,\eqref{eq:new_epsilon}, one can derive cosmological constraints on the three-body decay $N_2\rightarrow \bar N_1 \nu\nu$ from a hypothetical two-body decay characterized by a fictitious mass splitting $\epsilon'$ given by
\begin{equation}
	\epsilon'(\epsilon) \simeq \sqrt{\frac{13}{21}}\epsilon 
    \label{eq:mapping_mm}\,.
\end{equation}

\begin{figure}[t]
    \centering
    \includegraphics[width=0.6 \textwidth ]{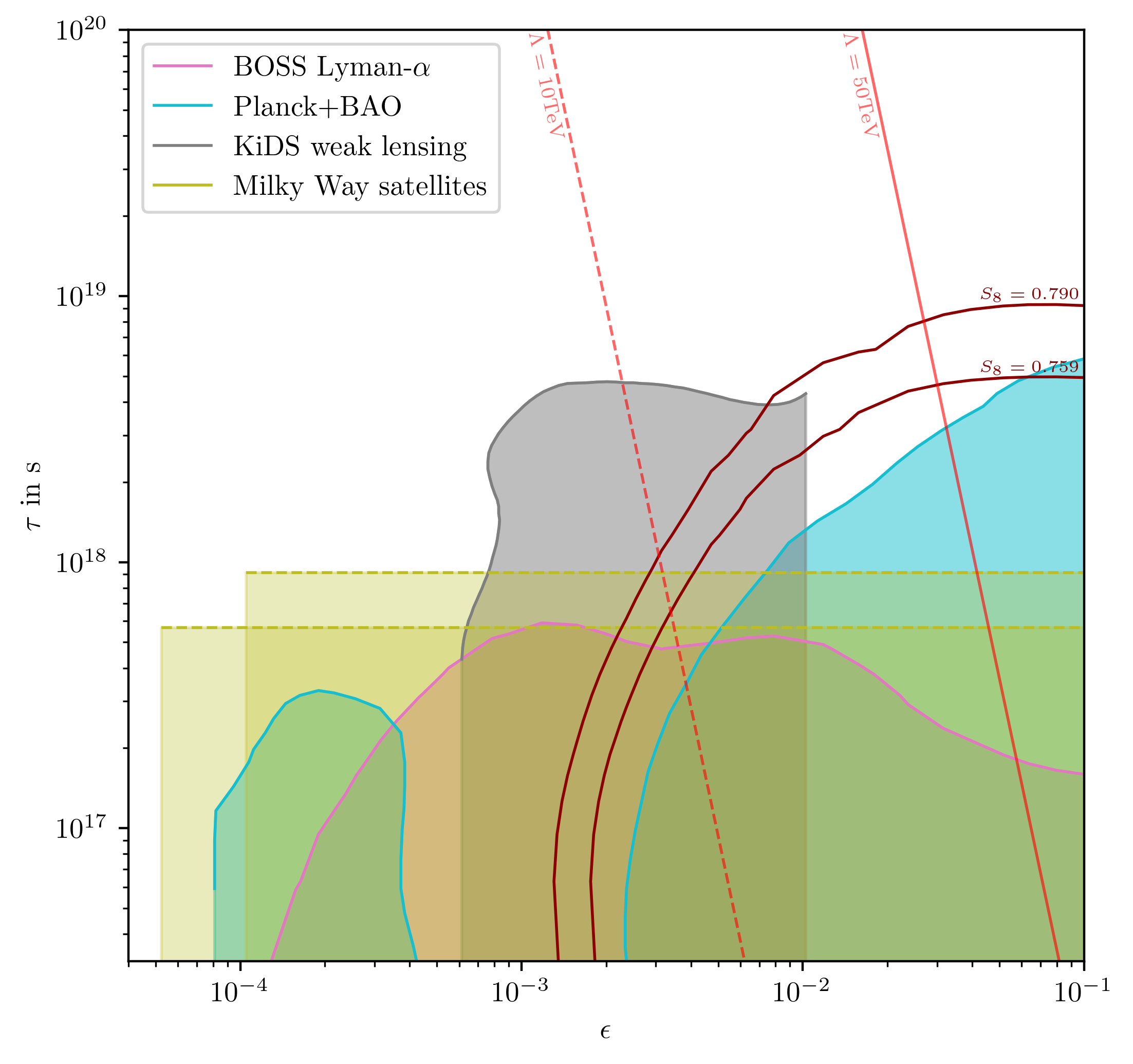}
    \caption{Same as Fig.~\ref{fig:cosmo_constraints}, but for the minimal decaying cold dark matter scenario $\text{DCDM} \ \rightarrow \ \text{WDM} \ + \ \text{DR}_a+ \ \text{DR}_b$ described in Sec.~\ref{sec:minimalmodel}. The plot also shows contours of the suppression scale $\Lambda=10$\,TeV and $50$\,TeV  of the dimension-eight operator Eq.\,\eqref{eq:Lagrangian} inducing the decay, assuming $M=0.3$\,GeV.
    }
    \label{fig:cosmo_constraints_3body}
\end{figure}

In Fig.\,\ref{fig:cosmo_constraints_3body} we show the cosmological constraints adapted to three-body decay and regions of the parameter space spanned by $\epsilon$ and $\tau$ that address the $S_8$ tension.
For illustration, we also show contour lines
in  Fig.~\ref{fig:cosmo_constraints_3body} of the suppression scale of the effective interaction operator Eq.\,\eqref{eq:Lagrangian}, at $\Lambda=10$\,TeV and $50$\,TeV, respectively. It would be interesting to investigate possible probes of the new particle species mediating the interaction of DM with neutrinos that are expected to exist at these scales.
However, in this work we focus on the signatures of the decay itself while remaining agnostic about the origin of the effective interaction.

\section{Constraints from the diffuse neutrino flux}
\label{sec:NeutrinoConstraints}

\begin{figure}[t]
    \centering
    \includegraphics[width=0.7\textwidth]{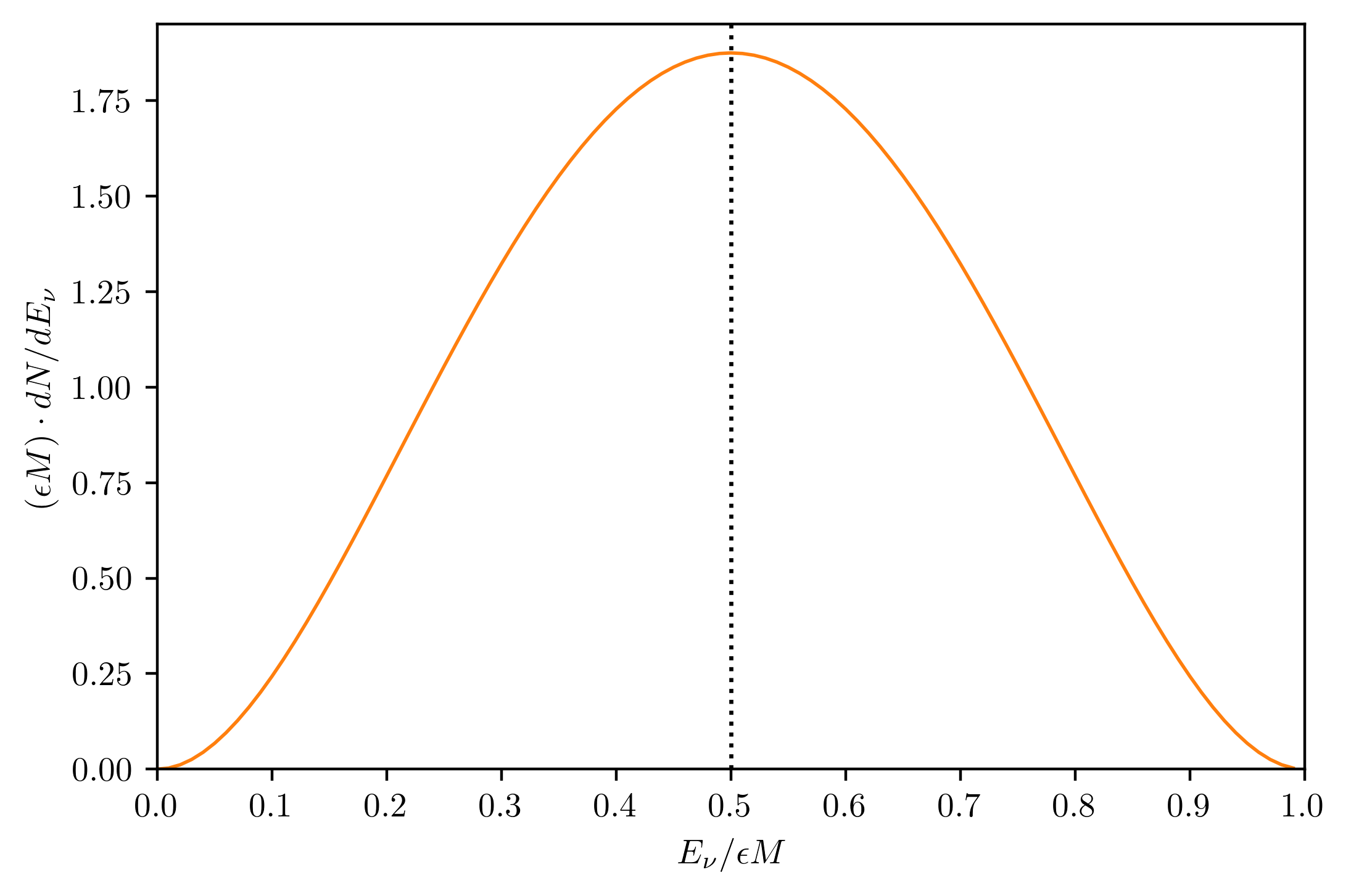}
    \caption{Neutrino spectrum $\dx N/ \dx E_\nu$ produced by three-body decays $N_2\to \bar N_1\nu\nu$ for $\epsilon\ll 1$, as a function of the neutrino energy normalized to its maximum value $\epsilon M$.
    }
    \label{fig:NeutrinoSpectrum}
\end{figure}

The model discussed in Sec.~\ref{sec:minimalmodel} produces 
a diffuse neutrino flux through the decays  $N_2\to\bar N_1\nu\nu$ as well as $\bar N_2\to N_1\bar\nu\bar\nu$, thus providing a possible test of this solution of the $S_8$ tension. Assuming that the bulk of the DM today is still in the form of cold dark matter (which is justified for the relevant lifetimes), the neutrino flux approximately reads
\begin{equation}
    \frac{\dx \Phi_{\nu}}{\dx E_\nu} \simeq \frac{1}{4\pi}\frac{1}{\tau M}\frac{1}{3}\frac{\dx N}{\dx E_\nu} D(\Omega)\,,
    \label{eq:neutrinoflux}
\end{equation}
where $dN/dE_\nu$ is the neutrino spectrum produced per DCDM decay normalized to one, and which is given in our model by
\begin{equation}
    \frac{\dx N}{\dx E_\nu} = \frac{1}{\Gamma}\frac{\dx \Gamma}{\dx E_\nu} = \frac{30 E_\nu^2 (M \epsilon-E_\nu )^2}{M^5 \epsilon ^5}\,.
    \label{eq:neutrino_spectrum}
\end{equation}
The energy spectrum of the neutrinos is shown for illustration  in Fig.~\ref{fig:NeutrinoSpectrum} for $\epsilon\ll 1$.
Further, the factor $1/3$ accounts for the three neutrino flavors, assuming that DM either decays into each flavor with equal rate, or that neutrino oscillations eventually cause all flavours to appear equally.  Finally, the so-called $D$ factor is defined as an integral of the DM density over the line of sight $l$ in a given angular region in the sky
\begin{equation}
    D(\Omega) = \int \dx\Omega \int \rho(l) \dx l\,,
\end{equation}
and thus depends on the DM distribution. For concreteness, we adopt the value quoted in~\cite{Arguelles:2022nbl}, $D(\Omega)=2.65\cdot 10^{23}\text{GeV}/\text{cm}^2$, which corresponds to  a Navarro-Frenk-White (NFW) profile with slope parameter $\gamma=1.2$ and scale radius $r_s=20$\,kpc, and a DM density $\rho=0.4\text{GeV}/\text{cm}^3$ at a distance $R_0=8.1$\,kpc from the Galactic center.

\begin{figure}
    \centering
    \includegraphics[width=0.63\textwidth]{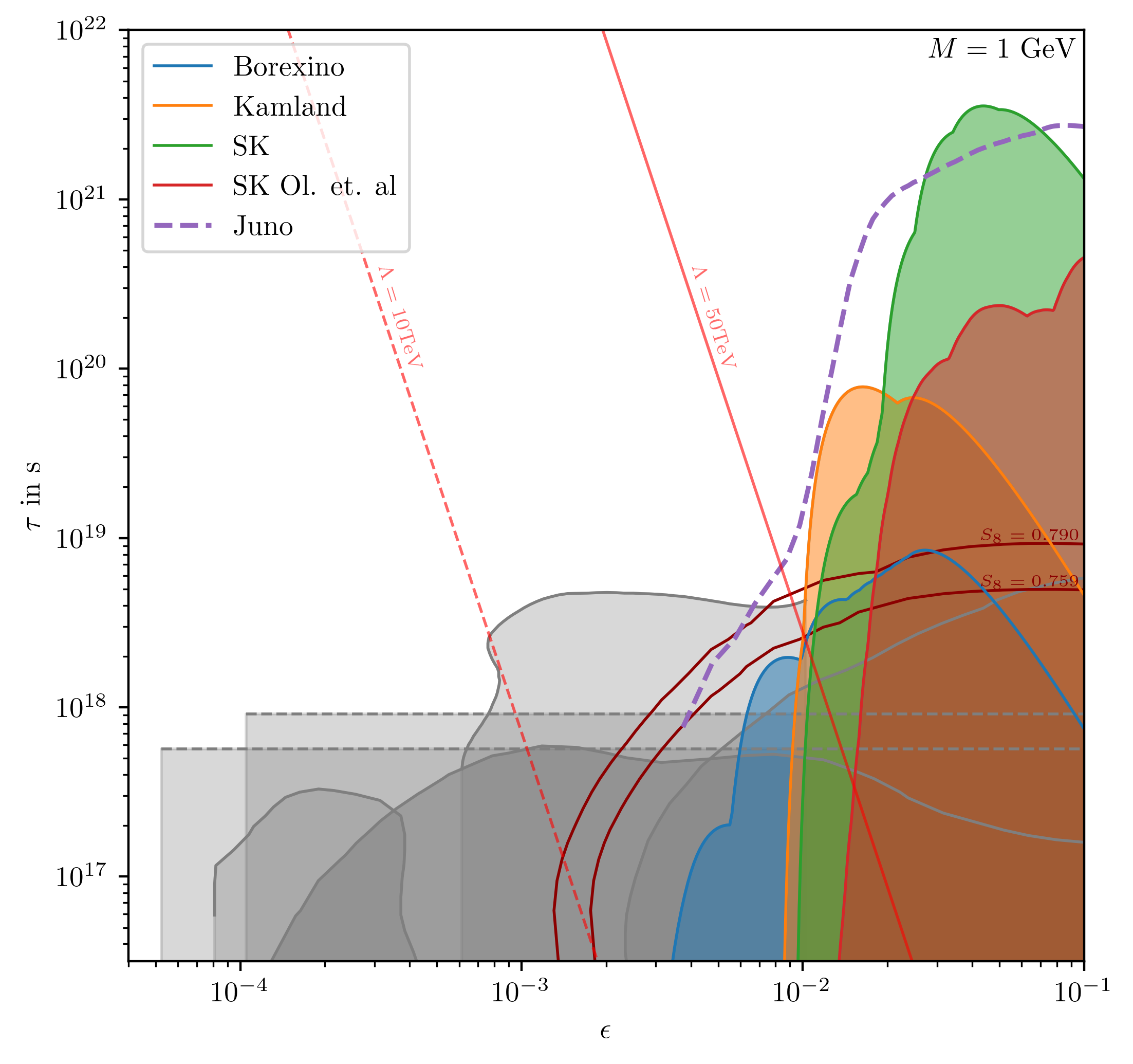}\\
    \includegraphics[width=0.63\textwidth]{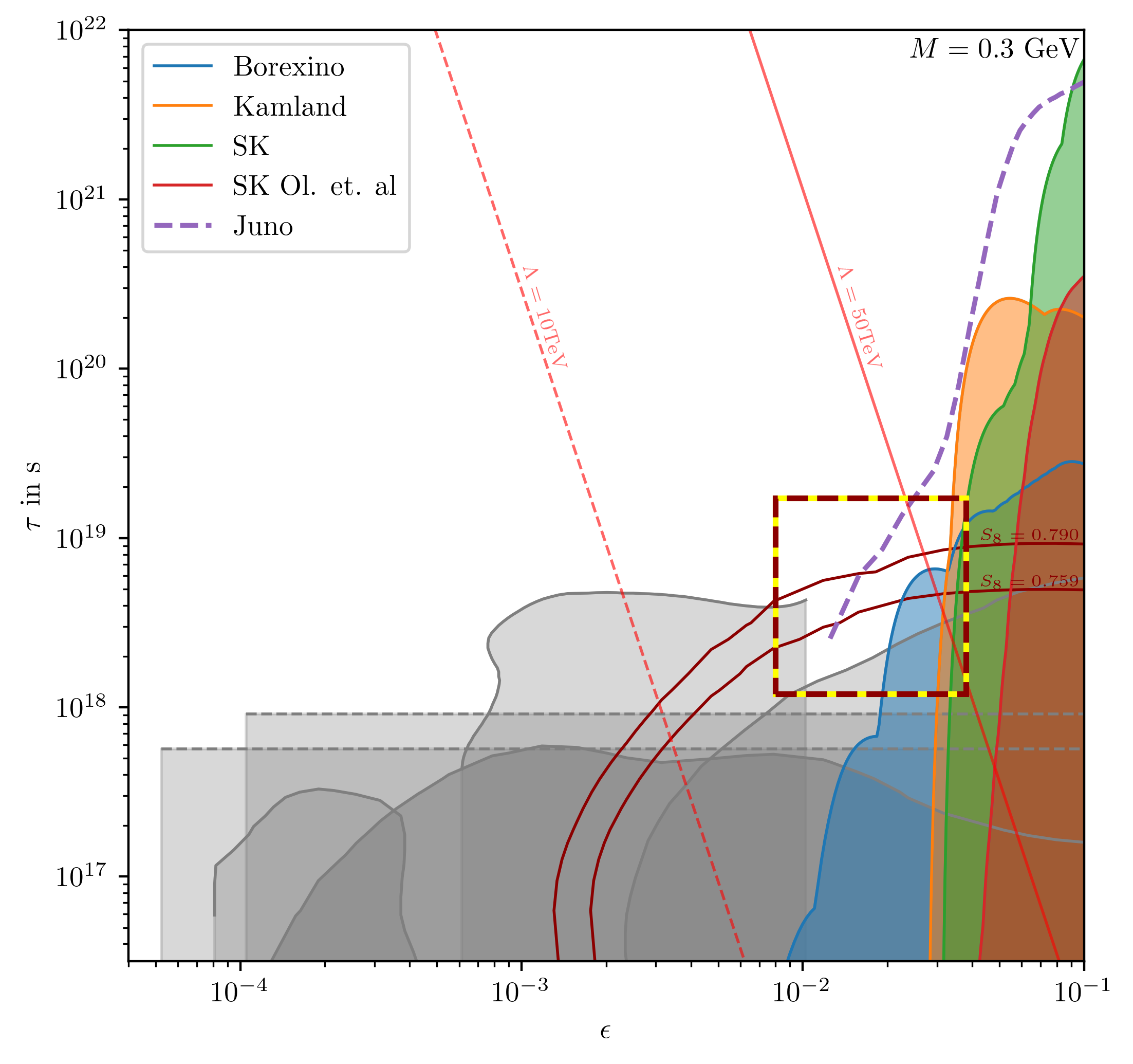}
    \caption{Allowed regions of the minimal decaying cold dark matter scenario $\text{DCDM} \ \rightarrow \ \text{WDM} \ + \ \text{DR}_a+ \ \text{DR}_b$ described in Sec.~\ref{sec:minimalmodel}, for $M=1$\,GeV (upper panel) and $M=0.3$\,GeV (lower panel). The gray regions are excluded by cosmological observations (see Sec.~\ref{sec:minimalmodel}), and the colored regions are excluded by neutrino experiments (see Sec.~\ref{sec:NeutrinoConstraints}). The red thick lines indicate the region of parameter space that can address the $S_8$ tension, while thin lines show contours of the suppression scale $\Lambda=10$\,TeV and $50$\,TeV  of the dimension-eight operator Eq.\,\eqref{eq:Lagrangian} inducing the decay.}
    \label{fig:NeutrinoConstraints}
\end{figure}

To the best of our knowledge, there are no published limits on the dark matter lifetime for this three-body decay spectrum (for other decay channels, see {\it e.g}~\cite{Arguelles:2022nbl}). We conservatively derive upper limits on the lifetime requiring that the flux generated in the decay does not exceed the measured flux. Concretely, we use the electron anti-neutrino flux measurements between $1.8$ and $16.8$\,MeV from Borexino~\cite{Borexino:2019wln},  between $8.3$ and $30.8$\,MeV from KamLAND~\cite{KamLAND:2021gvi}, as well as between $9.3$ and $34.3$\,MeV  from Super-Kamiokande (SK)~\cite{Super-Kamiokande:2021jaq}. Additionally, we recast SK limits for annihilating dark matter~\cite{Olivares-DelCampo:2017feq} for $10-200$\,MeV, and translate them into decay limits (indicated by SK Ol. et. al. in Fig.~\ref{fig:NeutrinoConstraints}).

We show in Fig.~\ref{fig:NeutrinoConstraints} as shaded colored regions the constraints from neutrino flux measurements on DCDM within the parameter space spanned by $(\epsilon,\tau)$, for two values of the DM mass: $M=1$\,GeV (top) and $M=0.3$\,GeV (bottom). We also show as thick red lines the values of parameters that could solve the $S_8$ tension.
We find that neutrino experiments for these mass scales constrain values $\epsilon\gtrsim {\cal O}(10^{-2})$, corresponding to neutrino energies above $\simeq 10$\,MeV, where neutrino detectors are most sensitive. The constraints from neutrino experiments are nicely complementary to the constraints from cosmology.
Interestingly, we find an allowed window in parameter space for low masses $M\lesssim$\,GeV, where low values of $S_8\lesssim 0.8$, as preferred by various cosmological data sets, are allowed both by neutrino and cosmological constraints (see highlighted box in the lower panel of Fig.~\ref{fig:NeutrinoConstraints}). 

Future neutrino experiments like JUNO~\cite{ColomerMolla:2023ppf,Akita:2022lit}, DUNE~\cite{DUNE:2020ypp,Arguelles:2019ouk} or Hyper-Kamiokande~\cite{Ruggeri:2023btm,Bell:2020rkw} will close in on the parameter space of the model. Specifically, we show in Fig.~\ref{fig:NeutrinoConstraints} the projected sensitivity of JUNO to the model, recasting the sensitivity of JUNO to the decay $\chi\rightarrow \nu \bar\nu$ given in~\cite{ColomerMolla:2023ppf}, and will explore regions of the parameter space that address the $S_8$ tension and which are allowed by current experiments.

\section{Dark matter production via freeze-in}
\label{sec:Production}

The effective operator Eq.\,\eqref{eq:Lagrangian} leading to the dark matter decay also leads to the production of dark matter particles via the processes $\nu\nu\to N_1 N_2$ and $\bar\nu\bar\nu \to\bar N_1\bar N_2$ when the temperature of the Universe is $T_{\text{EW}}\gtrsim T\gtrsim M$ (here, $T_{\text{EW}}\simeq 160$\,GeV is the temperature scale of electroweak symmetry breaking~\cite{DOnofrio:2015gop}). The values of the suppression scale favored by the DCDM solution to the $S_8$ tension  are $\Lambda={\cal O}(\text{TeV})$, which implies that the production process is very slow, and that the inverse annihilation processes have a negligible rate in the Universe (for details, see App.\,\ref{app:lowmass}). Therefore, within the DCDM scenario considered here, dark matter could be produced via the freeze-in mechanism.

The evolution of the total dark matter yield $Y$, defined as $Y=n/s$, with $n$  being the sum of the number densities $N_1$ and $N_2$ and $s$ the entropy density, is given by~\cite{Hall:2009bx, Edsjo:1997bg}
\begin{equation}
    \frac{\dx Y}{\dx x} = \frac{1}{\sqrt{g_{\text{eff}}}h_{\text{eff}}} \sqrt{\frac{5}{\pi}}\frac{135 M_{\text{pl}}}{4\pi^3M} x^4\frac{\gamma_{N_1N_2}}{M^4}\,,
    \label{eq:Y}
\end{equation}
where $x = M/T$ and $\gamma_{N_1N_2}$ is the DM production rate.
The abundance of $\bar N_1$ and $\bar N_2$ produced via the corresponding charge-conjugated process is equal, such that the total DM yield is given by $2Y$. Furthermore, we checked that conversion processes among the two species can be neglected for freeze-in (see App.~\ref{app:freeze-in}). 
We calculate $\gamma_{N_1N_2}$ by performing the phase-space integration over the squared matrix element and over the neutrino distribution (see App.~\ref{app:freeze-in} for details). 
Explicitly, it reads
\begin{equation}\label{eq:gammaN1N2}
    \gamma_{N_1N_2} =  \frac{\vew^4 M^8}{256 \pi^5 \Lambda^8} \frac{1}{x^8}\left(x^6 K_1\left(x\right)^2 + 2x^5K_1\left(x\right)K_2\left(x\right) + (4+x^2)x^4 K_2\left(x\right)^2\right)
    \,,
\end{equation}
with $K_1$ and $K_2$ modified Bessel functions of the second kind of order one and two, respectively.  
Notice the strong increase with temperature with $\gamma_{N_1N_2}\propto T^8$ in the limit of $T\gg M$.  This can be related to the increase of the cross section with center-of-mass energy, which is in turn related to the fact that the interaction is described by an effective four-fermion vertex.

We have solved the Boltzmann equation, and we have determined the yield at $x\rightarrow\infty$, $Y_\infty$. As initial condition, 
we assume an instant reheating of the universe at a temperature $T_\text{rh}\leq T_\text{EW}$, at which the dark matter yield is equal to zero. Relaxing the assumption of instant reheating could lead to additional contributions to the DM abundance, depending on the specific reheating model~\cite{Giudice:2000ex}. Finally, we calculate the total DM abundance today accounting for the equal yields of $N_1$, $\bar N_1$, $N_2$ and $\bar N_2$ from
\begin{equation}
    \Omega_{\text{dm}} h^2
    \simeq \frac{2 M Y_\infty s_0}{\rho_{\text{crit},0}/h^2} \approx \frac{2.74 \cdot 10^8 \cdot Y_\infty \cdot 2 M}{\text{GeV}}\,,
\end{equation}
where $\rho_{\text{crit},0}$ and $s_0$ are the critical and entropy densities today, respectively. 

\begin{figure}
    \centering
    \includegraphics[width=0.6\textwidth]{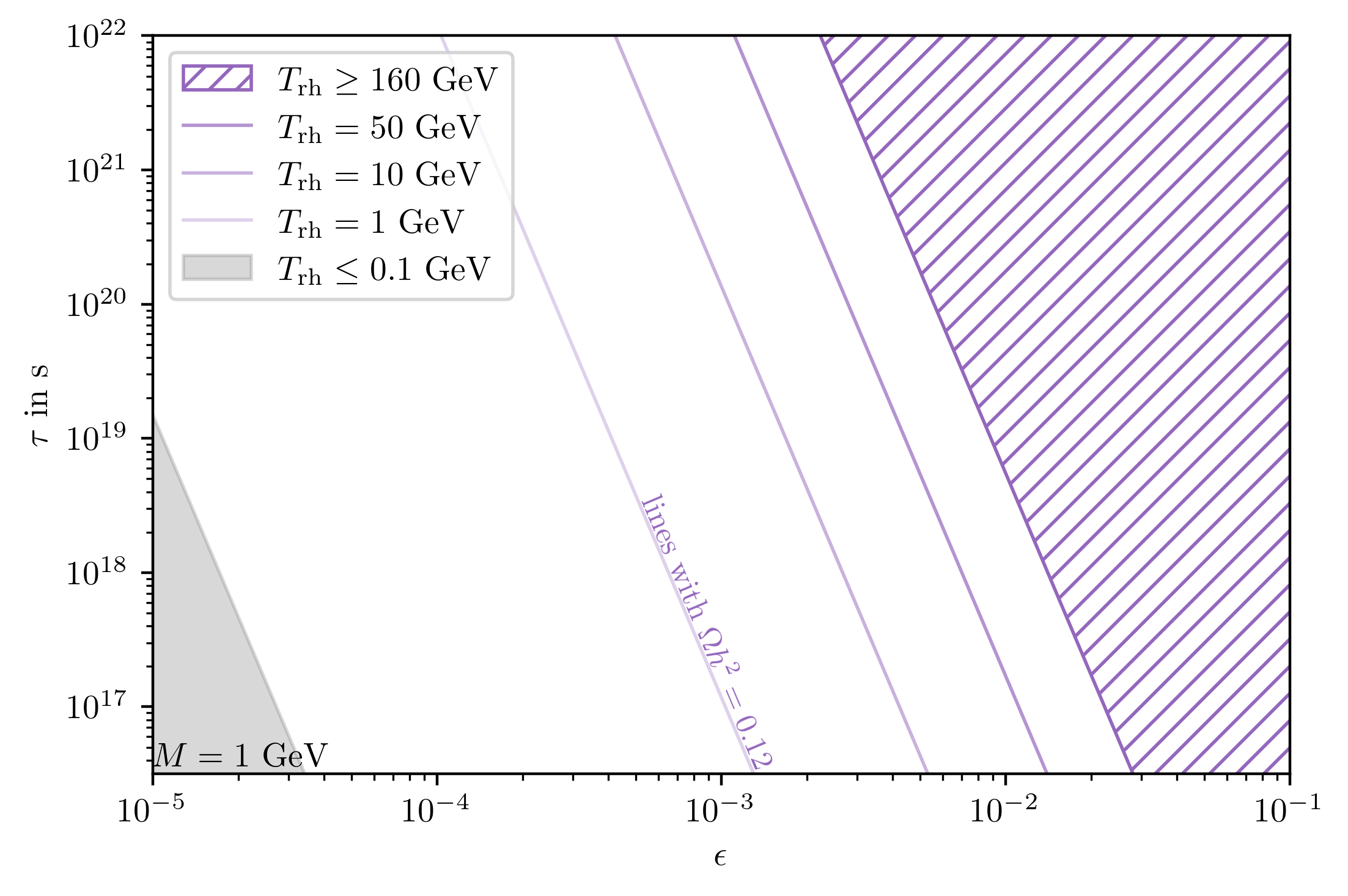}
    \caption{Values of lifetime and relative mass splitting for which the measured DM abundance $\Omega h^2=0.12$ can be reproduced via freeze-in, for $M=1$\,GeV and for different values of the reheating temperature below $T_{\text{EW}}\simeq 160$\,GeV, due to the scatterings $\nu\nu\to N_1 N_2$ and $\bar\nu\bar\nu \to\bar N_1\bar N_2$ induced by the operator 
    Eq.\,\eqref{eq:Lagrangian}. In the hatched region the production is suppressed due to the large value of $\Lambda$, and in the grey region the production is suppressed due to the small phase space available.
    }
    \label{fig:production}
\end{figure}

\begin{figure}[t]
    \centering
    \includegraphics[width=0.6\textwidth]{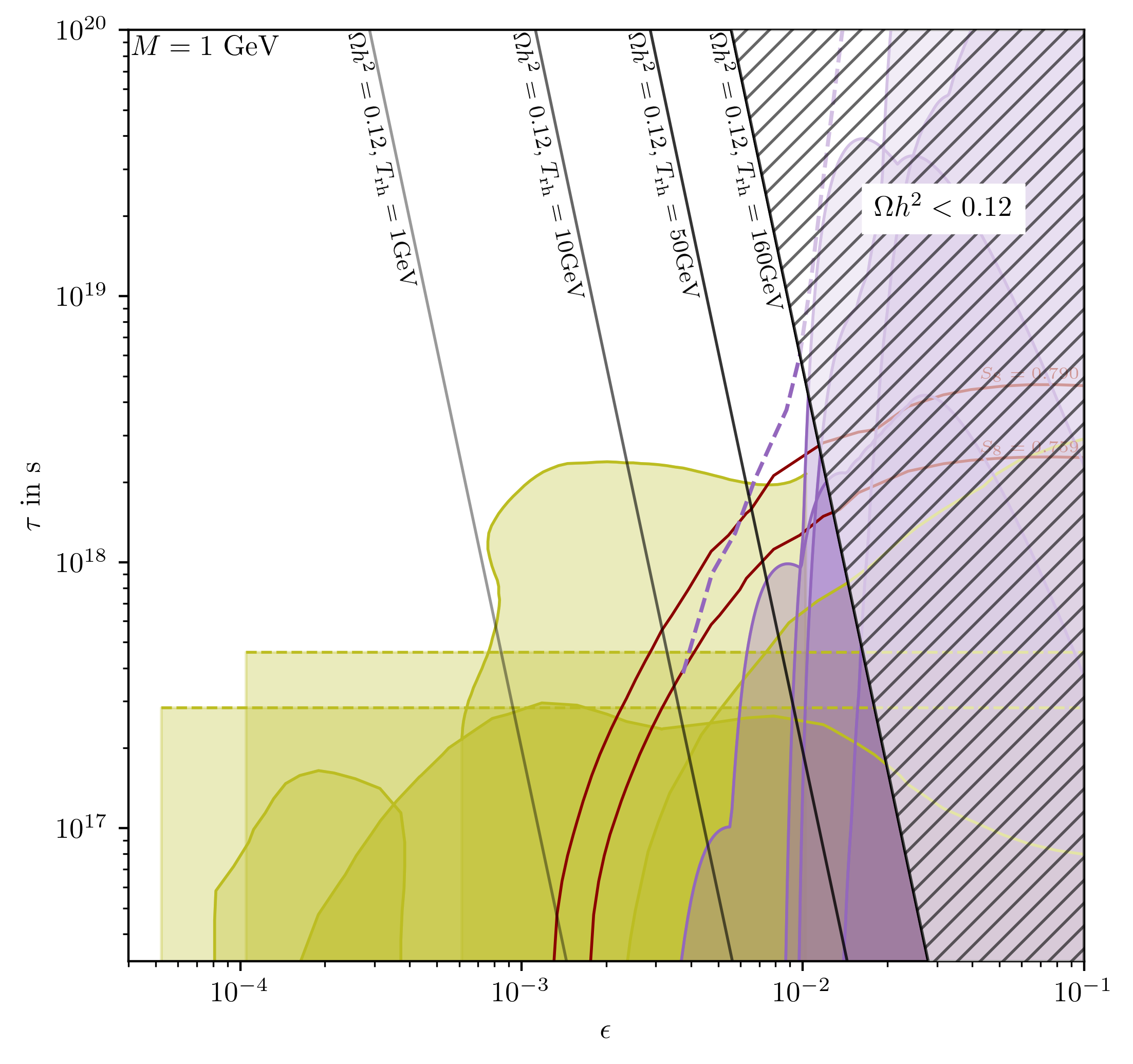}
    \includegraphics[width=0.6\textwidth]{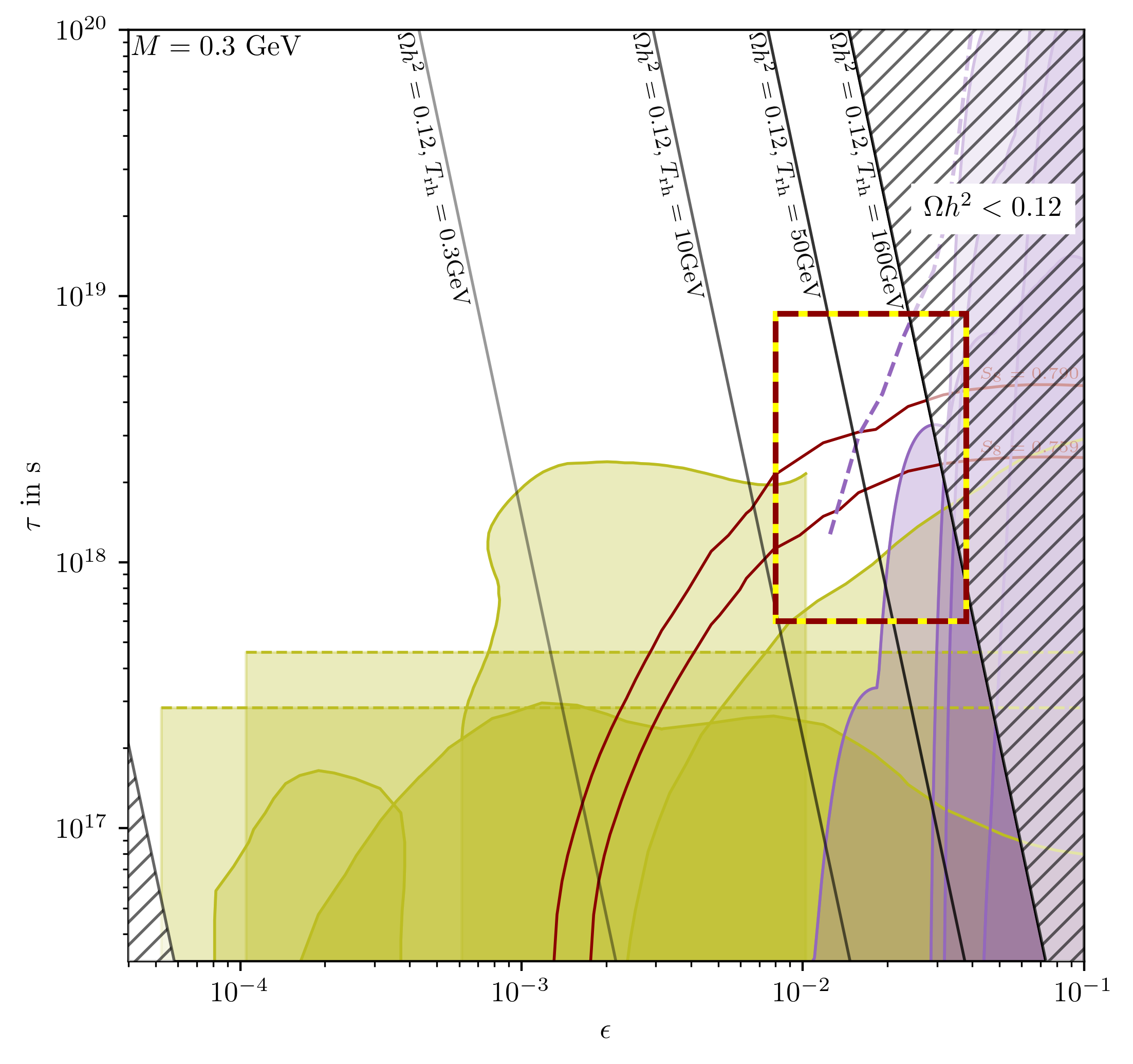}
    \caption{Recasting of Fig.~\ref{fig:NeutrinoConstraints},  for the scenario where DM is produced via freeze-in.
   The black lines show contours for which $\Omega h^2=0.12$ for various reheating temperatures. In the hatched regions, the DM is underproduced ({\it cf.}~Fig.~\ref{fig:production}). 
   The box in the lower panel highlights the regions in parameter space where DCDM can address the $S_8$ tension and is compatible with all cosmological and experimental constraints, along with freeze-in production.}
    \label{fig:exclusion_production_1GeV}
\end{figure}

In Fig.\,\ref{fig:production}, we show the regions in parameter space $(\epsilon,\tau)$ for which freeze-in production matches the value $\Omega h^2=0.12$ preferred by Planck~\cite{Planck:2018vyg},  for $M=1$\,GeV and for various values of the reheating temperature $T_{\text{rh}}$. As discussed above we limit ourselves to the regime $T_{\text{rh}}<T_{\text{EW}}$, which is realized below the hatched region. 
Efficient freeze-in production is possible only for $T_{\text{rh}}\gtrsim M$, since otherwise the typical thermal energy of neutrinos is not sufficient to produce DM particles. Specifically, for $T<M$ the Bessel functions entering the production rate Eq.\,\eqref{eq:gammaN1N2} feature an exponential Boltzmann suppression and thus we additionally exclude the region where $T_{\text{rh}}<0.1 M$ is required to reach the relic abundance in gray.
Notably, the values preferred by freeze-in production are in the ballpark of the  the values that address  the $S_8$ tension,  $\tau\simeq 10^2$\,Gyrs and $\epsilon\simeq 10^{-2}$.

Production via freeze-in leads to an initial population of DM for which only 50\% are in the form of the heavier state $N_2+\bar N_2$ while 50\% are already produced in the lighter state $N_1+\bar N_1$. Thus, we have to generalize the results of our previous analyses with $100\%$ DCDM. This can be easily done in the limit of $\tau \gg t_0$, which is always the case in the parameter space of interest, and we show in App.\,\ref{app:freeze-in} that the cosmological and astrophysical constraints derived previously can be mapped on the freeze-in scenario by simply re-scaling the lifetime by a factor of two.

Using this mapping, we overlay cosmological and astrophysical constraints with the requirement from producing the observed DM abundance via freeze-in for various viable reheating temperatures in Fig.\,\ref{fig:exclusion_production_1GeV}. We note that for DM masses of order GeV, the region in parameter space relevant for the $S_8$ tension and compatible with constraints from neutrino flux measurements as well as CMB data is also compatible with freeze-in production. This is remarkable, since all interactions are generated by a single effective operator Eq.\,\eqref{eq:Lagrangian} within the minimal model considered in this work. It is also interesting to note that this region of parameter space can be tested by upcoming neutrino experiments such as JUNO (dashed lines in Fig.\,\ref{fig:exclusion_production_1GeV}) as well as future weak lensing surveys sensitive to $S_8$.

\section{Other possible signatures}
\label{sec:other_signatures}

In this section we discuss potential additional signatures that necessarily arise from the interaction described by the effective operator Eq.\,\eqref{eq:Lagrangian}.

\subsection{Dark matter decay into charged particles}
\label{sec:decay_charged_particles}

\begin{figure}[t]
    \centering
    \includegraphics[width=0.3\textwidth]{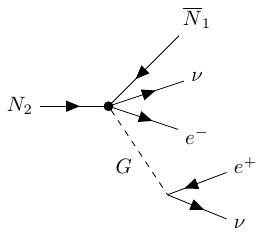}
    \includegraphics[width=0.33\textwidth]{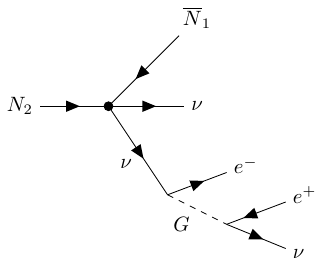}
    \includegraphics[width=0.33\textwidth]{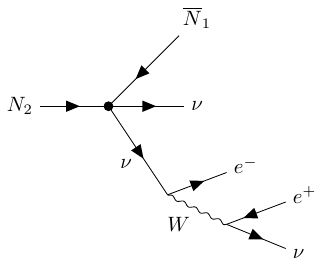}
    \caption{Decay producing an additional pair of charged particles via Goldstone and $W$ boson processes. In unitary gauge, only the last diagram  contributes, while in general only their sum is gauge independent.}
    \label{fig:decaychargedparticles}
\end{figure}

As discussed in Sec.~\ref{sec:minimalmodel}, DM decays into visible particles need to be strongly suppressed in order to satisfy positron and gamma-ray flux limits. Due to lepton number and charge conversation, the simplest form of such a decay is $N_2\rightarrow \bar{N_1}\nu e^- e^+ \nu$. This process can be mediated by either the Goldstone, arising from the Higgs-doublet, or the $W$ boson and we show all contributing diagrams in Fig.~\ref{fig:decaychargedparticles}. Note that in general only their sum is gauge independent (see App.~\ref{app:decay_charged_particles}).

Performing the five-body phase-space integration analytically in the limit $\epsilon\ll 1$, the resulting decay width reads
\begin{equation}
    \Gamma_{N_2\rightarrow\bar{N}_1 \nu \nu e^+ e^-} = \frac{v_\text
 {EW}^4 g^4}{16 m_\text{W}^4}\frac{(\epsilon M)^9}{7741440\pi^7 \Lambda^8}\,.
 \label{eq:5bodydecay_result}
\end{equation}
The scaling with $\epsilon^9$ results from the additional phase-space suppression which scales as $\epsilon^7$ while the squared matrix element scales as $\epsilon^2$ as for the three-body decay (see App.~\ref{app:decay_charged_particles} for details).
Using Eq.\,\eqref{eq:maindecaywidth}, we find that the branching fraction is
\begin{equation}\label{eq:branchingratio}
    \frac{\Gamma_{N_2\rightarrow\bar{N}_1 \nu \nu e^+ e^-}}{\Gamma_{N_2\rightarrow\bar{N}_1 \nu \nu}} = \frac{(\epsilon M)^4}{6048\pi^4 \vew^4}\
    \approx 5\cdot 10^{-28} \left(\frac{\epsilon M}{\text{MeV}}\right)^4\,.
\end{equation}
We then find that the DCDM solution to the $S_8$ tension implies  partial decay rates into electrons of order $\simeq 10^{-41}\s^{-1}$, which is far away from the sensitivity of current searches for cosmic electrons/positrons in the MeV range, $\Gamma_{e^+}\lesssim 10^{-27}-10^{-29}\s^{-1}$~\cite{DelaTorreLuque:2023olp,DelaTorreLuque:2023cef}.

\subsection{Dark matter decay into photons}
\label{sec:decay_photons}

\begin{figure}[t]
    \centering
    \includegraphics[height=0.18\textheight]{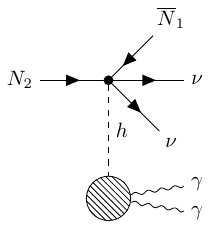} \hfill
    \includegraphics[height=0.15\textheight]{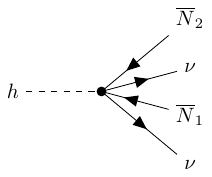} \hfill
    \includegraphics[height=0.15\textheight]{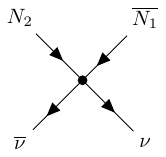}
    \caption{Diagrams leading to possible additional signatures of the DCDM model: dark matter decay with photons in the final state (left panel), Higgs invisible decay channel (middle panel), and neutrino-DM scattering (right panel).
    }
    \label{fig:feynmandiagrams_othersignatures}
\end{figure}

The interaction Lagrangian Eq.\,\eqref{eq:Lagrangian} leads to an effective operator of the form
\begin{equation}
	\mathcal{L}_{\text{eff}} \supset \frac{\vew}{2\Lambda^4}\,h \,\bar{\nu} P_R N_2 \, \bar{\nu} P_R N_1 + \text{h.c.}\,,
    \label{eq:Lagrangian_higgs}
\end{equation}
which differs from Eq.\,\eqref{eq:nunuNN} in the substitution of the
Higgs vacuum expectation value by a Higgs boson. The Higgs boson is too heavy to be produced on-shell, but it could decay off-shell into two photons as shown in Fig.~\ref{fig:feynmandiagrams_othersignatures} (left panel).
The branching ratio reads 
\begin{align}\label{eq:branchingratioHiggs}
    \frac{\Gamma_{N_2\to\bar N_1\nu\nu\gamma\gamma}}{\Gamma_{N_2\to\bar N_1\nu\nu}} \simeq \frac{\alpha_{\text{em}}^2}{ \pi^2}\frac{(\epsilon M)^8}{m_h^4 v_{EW}^4}\,,
\end{align}
see App.\,\ref{app:decay_charged_particles} for details.
For the relevant parameter space with $\epsilon\simeq{\cal O}(10^{-2})$ and DCDM masses in the GeV range the branching ratio is thus very strongly suppressed, of the order of $10^{-39}$.  For the typical DCDM lifetimes required to address the $S_8$ tension, the expected width into photons is  the order of $\Gamma_\gamma\simeq 10^{-39}\,{\rm s}^{-1}$, which is again far away from the sensitivity of current experiments, $\Gamma_\gamma\lesssim 10^{-23}-10^{-30}\,{\rm s}^{-1}$~\cite{Berteaud:2022tws,Calore:2022pks,Essig:2013goa}. The four-body decay  $N_2\to\bar N_1\nu\nu\gamma$ associated to the five-body decay $N_2\rightarrow \bar{N_1}\nu e^- e^+ \nu$ upon closing the electron/positron lines in a loop is also expected to have a width well below the current sensitivity of experiments.

\subsection{Invisible Higgs decay}
\label{sec:invisible_higgs}

The effective interaction Eq.\,\eqref{eq:Lagrangian_higgs} derived from the
interaction  Lagrangian  Eq.\,\eqref{eq:Lagrangian} also implies a novel decay mode of the Higgs particle into DM particles and neutrinos, as shown in the middle of Fig.\,\ref{fig:feynmandiagrams_othersignatures}.

The decay width reads
\begin{equation}
    \Gamma_{\text{h}}^{\text{inv}} = \frac{1}{4 m_h} \frac{\vew^2}{30 \pi^5 \Lambda^8}\left(\frac{m_h}{4}\right)^8 \approx 1.37 \cdot 10^{-20}\text{MeV} \,\left(\frac{\text{MeV}}{\epsilon M}\right)^5 \left(\frac{100 \gyr}{\tau}\right)\,,
\end{equation}
which has been normalized to the typical values of $\epsilon M$ and $\tau$ that address the $S_8$ tension. In view of the value of the Higgs decay width into SM final states, $\Gamma_{\text{h}}^{\text{SM}}\simeq 3.2$\,MeV, and the current upper limit on the invisible decay width of the Higgs, ${\rm BR}_{h\rightarrow {\rm inv}}\lesssim 12\%$~\cite{ParticleDataGroup:2022pth},  one concludes that this invisible decay is far below the current experimental sensitivity when $M\gtrsim 10$\,MeV.  
For lower masses, we refer to a more detailed discussion in App.~\ref{app:lowmass}.

\subsection{Neutrino-DM scattering}
\label{sec:neutrinoDM_scattering}

The effective interaction Eq.\,\eqref{eq:nunuNN} responsible for DCDM decay necessarily also gives rise to scattering between the DM particles and neutrinos, as shown in Fig.\,\ref{fig:feynmandiagrams_othersignatures} on the right.
More concretely, the neutrino can induce an exothermic or endothermic scattering with $N_2$ or $N_1$ respectively. 
These interactions can lead to a coupled neutrino-DM fluid exhibiting pressure such that DM fluctuations are damped on small scales, or even undergo dark acoustic oscillations. Thus, a neutrino-DM interaction can lead to observable deviations in the CMB, LSS, Lyman-$\alpha$ forest, and subhalo counts. In the literature, the case of an elastic neutrino-DM scattering has been  extensively studied, leading to constraints on the cross section $\sigma_{\nu\text{DM}}$, see \emph{e.g.}~\cite{DiValentino:2017oaw,Brax:2023tvn,Giare:2023qqn,Wilkinson:2014ksa,Diacoumis:2018ezi,Mosbech:2020ahp,Escudero:2015yka,Hooper:2021rjc,Olivares-DelCampo:2017feq,Akita:2023yga,Mangano:2006mp,Diamanti:2012tg}.

To estimate whether neutrino free-streaming is altered by our effective operator, we compare the scattering rate to the Hubble rate. As shown in  App.\,\ref{app:NuDM} the scattering rate in the thermal plasma around recombination scales as $\Gamma_\nu \propto T^4$, while the Hubble rate only scales as $H \propto T^2$.
The strong temperature dependence implies that for temperatures relevant for cosmological limits, the scattering rate is severely suppressed, $\Gamma_\nu/H\ll 1$. Therefore, we do not expect any significant effect of the neutrino-DM scattering process in cosmology.

The  neutrino-DM scattering cross section could also be constrained from the observation of high-energy neutrinos from astrophysical sources in neutrino telescopes, which imply that the neutrino fluxes have not been significantly attenuated by interactions with DM during their propagation from the production point to the Earth. The detection of high-energy neutrinos from the blazar TXS0506+056 by IceCube~\cite{IceCube:2018cha,Tanaka2017FermiLATDO} allows to set constraints on the neutrino-DM cross section at  $E_\nu \simeq290$\,TeV. Conservative limits have been derived in~\cite{Choi:2019ixb}, and have been improved including the effect of the dark matter spike around the central black hole of the blazar in~\cite{Ferrer:2022kei,Cline:2023tkp}. 

At very high neutrino energies, the cross section of the scattering process $N_2 \bar\nu \rightarrow \bar{N_1}\nu$ reads 
\begin{align}\label{eq:DMneutrinoscattering_approx}
    \sigma_{N_2 \bar\nu \rightarrow \bar{N_1}\nu} &\simeq\frac{v_{EW}^4 M E_\nu}{128 \pi \Lambda^8}\simeq  2\times 10^{-49} \text{cm}^2 \left(\frac{E_\nu}{290 \text{TeV}}\right) \left(\frac{\text{MeV}}{\epsilon M}\right)^4 \left(\frac{10^{-2}}{\epsilon}\right) \left(\frac{100 \text{Gyrs}}{\tau}\right)\,,
\end{align}
and similarly for $N_1 \bar\nu \rightarrow \bar{N_2}\nu$
(see App.\,\ref{app:NuDM} for details). For typical parameters addressing the $S_8$ tension and DM masses above ${\cal O}(\text{MeV})$ this is far below the upper limit derived from attenuation of high-energy neutrinos emitted from the blazar TXS0506+056 which are of order $10^{-29}$cm$^2$~\cite{Ferrer:2022kei,Cline:2023tkp}.
Note that despite the large energy of the IceCube neutrinos, the validity condition $\sqrt{s}\ll\Lambda$ is satisfied for the relevant parameter values.
The possibility of very low mass DM in the MeV range is discussed in App.\,\ref{app:lowmass}.

Further constraints on the neutrino-DM scattering strength can be derived from the neutrinos detected from the supernova 1987A~\cite{Raffelt:1996}, that probe neutrino energies at the MeV scale, \emph{i.e.}~intermediate between CMB and IceCube energy scales. The upper limits of order $10^{-23}$cm$^2$ as quoted in~\cite{Ferrer:2022kei} are however somewhat weaker. Similarly, constraints can be derived in the same energy range from diffuse supernovae neutrinos affecting the DM density profile, as done recently in~\cite{Heston:2024ljf}. However, these also produce weaker limits for our model.

\section{Conclusion and outlook}
\label{sec:Conclusion}

We have constructed and investigated a minimal model of decaying dark matter that can address the $S_8$ tension. The model consists of a cold dark matter particle that decays into a warm dark matter particle, quasi-degenerate in mass with the former, and two particles of ``dark radiation'', that we identify with the Standard Model neutrinos. This leads to a mild suppression of the matter power spectrum on small scales and at late times, in qualitative agreement with the $S_8$ measurements, if the cold dark matter lifetime is  $\tau \simeq \mathcal{O}(10-100)$\,Gyrs and the relative mass splitting between the cold and the warm dark matter particles is  $\epsilon \simeq 10^{-2}- 10^{-3}$. In our model, both cold and warm dark matter candidates carry lepton number, and have opposite charges under a new global $U(1)$ symmetry. This assignment allows the decay of the cold dark matter into warm dark matter and two neutrinos via a higher dimensional operator. On the other hand, the  decays into charged leptons or into photons are very suppressed, thus evading the stringent limits on the lifetime of the decaying dark matter from cosmic electron/positron and gamma-ray experiments.

We have found that neutrino experiments  like Borexino, KamLAND  and Super-Kamiokande could detect signals of the diffuse neutrino flux generated in the decay, and that the upcoming JUNO neutrino observatory has the potential to probe the regions of parameter space addressing the $S_8$ tension if the DM mass is below $\simeq 1$\,GeV. Furthermore, the same interaction that mediates the three-body dark matter decay can also explain the abundance of dark matter by the freeze-in mechanism, via the process of conversion of two neutrinos into two dark matter particles in the Early Universe. 

Finally, we have also investigated other possible signatures of the model, including the indirect detection of electron/positrons or gamma-rays from the decay, the invisible Higgs decay width, neutrino-dark matter scattering in the Early Universe, or the attenuation of the high-energy neutrino flux from blazars. For dark matter mass in the GeV mass range these signals are too faint to be detected, although they might be observed if the dark matter mass is in the MeV mass range. It would be interesting to explore possible UV completions of our decaying dark matter scenario, and whether the new particles could lead to  additional signals in experiments. We leave this direction for future work.

\subsection*{Acknowledgements}

We acknowledge support by the DFG Collaborative Research Institution Neutrinos and Dark Matter in Astro- and Particle Physics (SFB 1258) and the Excellence Cluster ORIGINS - EXC-2094 - 390783311.

\appendix
\section{Five-body dark matter decays}
\label{app:decay_charged_particles}

\subsection{Decay channel \texorpdfstring{$N_2\rightarrow \bar{N}_1\nu e^- e^+ \nu$}{into charged particles}}

In this section we provide some details  for the $N_2\rightarrow \bar{N}_1\nu e^- e^+ \nu$ process discussed in Sec.~\ref{sec:decay_charged_particles}. Due to the Higgs-doublet $\tilde H=i\sigma_2 H^*=\left( (\vew + h - iG^0)/\sqrt{2}, -G^- \right)$, entering the effective interaction operator in Eq.\,\eqref{eq:Lagrangian}, DM can couple to an electron or positron via the longitudinal polarization of a $W$ boson represented by the Goldstone boson $G$, \emph{e.g.}
\begin{equation}
	\mathcal{L}_{\text{eff}} \supset \frac{\vew}{2\Lambda^4} \hspace{1.5mm} \bar{\nu} P_R N_2 \hspace{1.5mm} e^+ P_R N_1 G^- +\text{h.c.}\,.
    \label{eq:Lagrangian_Goldstone}
\end{equation}
The Goldstone boson can then in turn decay into another charged lepton and neutrino pair, giving rise to a decay channel $N_2\rightarrow \bar{N_1}\nu e^- e^+ \nu$. The same decay can also be generated from $N_2\rightarrow \bar{N_1}\nu \nu$, where additionally one neutrino emits a Goldstone or $W$ boson. All of these three contributions are shown in Fig.~\ref{fig:decaychargedparticles}, where the individual diagrams depend on the choice of gauge fixing in general, while the complete matrix element should be gauge-fixing independent. To verify this, we can write down the matrix elements for each contribution in a general $R_\xi$ gauge where the Goldstone propagator appearing in the first two diagrams takes the form $\frac{i}{\kappa_1^2 -\xi m^2_W}$. The $W$ boson propagator needed for the last diagram is given by
\begin{equation}
    \frac{-i}{\kappa_1^2 - m^2_W} \left[ g_{\mu\nu} - (1-\xi)\frac{\kappa_{1 \mu} \kappa_{1 \nu}}{\kappa_1^2 -\xi m^2_W} \right] = \frac{-i g_{\mu\nu}}{\kappa_1^2 - m^2_W}+ \frac{i \kappa_{1 \mu} \kappa_{1 \nu}}{\kappa_1^2-m^2_W} - \frac{i \kappa_{1 \mu} \kappa_{1 \nu}}{\kappa_1^2-\xi m^2_W}\,,
    \label{eq:Wpropagator}
\end{equation}
where we have rewritten it to recognise that only the last term depends on $\xi$ and resembles a Goldstone propagator. In turn, we find that this term is responsible for cancelling both contributions from the first and second Goldstone process on the amplitude level, which is to be expected since the dependence on $\xi$ needs to drop out to ensure gauge independence. 
Consequently, we can safely work in unitary gauge with $\xi \to \infty$ where the Goldstone contributions vanish naturally. Thus, the only left-over term now stems from the first two terms in Eq.\,\eqref{eq:Wpropagator} from the $W$ boson process and is given by
\begin{align}
	-i\mathcal{M} = \frac{1}{\Lambda^4}& \frac{1}{\kappa_2^2} [\bar{u}(k_2) \gamma_\mu P_L v(l_2)] \left[ \frac{m^2_W}{\kappa_1^2 -m^2_W} g^{\mu\nu} - \frac{\kappa_1^\mu \kappa_1^\nu}{\kappa_1^2-m^2_W}\right]  \nonumber\\
	\left( \right. &[\bar{u}(l_1) \gamma_\nu P_L \cancel{\kappa_2}P_R u(p_1)] [\bar{u}(k_1)P_R v(p_2)] \\
	+ & [\bar{u}(k_1)\gamma_\nu P_L \cancel{\kappa_2}P_R u(p_1)] [\bar{u}(l_1,t_1)P_R v(p_2)] \left.\right)\,. \nonumber
\end{align}
Hereby, we chose $p_{1,2}$ for the momentum of the $N_{2,1}$ particles, $l_{1,2}$ for the $e^-$ and $e^+$ and $k_{1,2}$ for the final state neutrinos. In the propagators, the $W$ boson momentum is indicated by $\kappa_1$ and the neutrino momentum by $\kappa_2=\kappa_1 + l_1$. Note that we do not include the exchange of both indistinguishable neutrinos $(k_1 \leftrightarrow k_2)$ here which would cause an interference term to appear in the squared matrix element. Technically, this is justified if the leptons have different flavour, \emph{e.g.}~for $N_2\to\bar{N}_1\nu_\mu e^-e^+\nu_e$, 
as generated from an effective interaction involving $L_e$ and $L_\mu$ lepton doublets. Since the resulting decay width is many orders of magnitudes below the relevant limits, we expect that this is also the case for $N_2\to\bar N_1\nu_e e^-e^+\nu_e$, provided the effective interaction strength is comparable for all flavour combinations. The squared matrix element with this simplification and the limit of $m_{\nu,e}\to 0$ and $m_W \to \infty$, then takes the form
\begin{align}
	\overline{|{\cal M}|^2} =& \frac{v_\text{EW}^4 g^4}{\Lambda^8 \kappa_2^4 m_\text{W}^4} \left( 4(k_2\cdot l_1)(k_1\cdot p_1)(\kappa_2\cdot l_2)(\kappa_2\cdot p_2) + 4(k_2\cdot l_1)(k_1\cdot p_2)(\kappa_2\cdot l_2)(\kappa_2\cdot p_1) \right. \nonumber\\
    & -2\kappa_2^2(k_2\cdot l_1)(k_1\cdot p_2)(l_2\cdot p_1) -2\kappa_2^2(k_2\cdot l_1)(k_1\cdot p_1)(l_2\cdot p_2) \nonumber \\
    & -2(p_1\cdot p_2)(\kappa_2\cdot k_1)(k_2\cdot l_1)(\kappa_2\cdot l_2) + \kappa_2^2(p_1\cdot p_2)(k_1\cdot l_2)(k_2\cdot l_1) \left. \right)\,,
    \label{eq:M2_chargedparticles}
\end{align}
where we made use of FeynCalc~\cite{Shtabovenko:2020gxv}. Note, that the $W$ boson mass drops out because of $m_\text{W}=v_\text{EW} g /2$, where $g$ is the $SU(2)_L$ gauge coupling constant.

For the decay width, we have to compute the 5-body phase-space integral $\dx\Phi_5$ over the matrix element
\begin{equation}
    \Gamma_{N_2\rightarrow\bar{N}_1 \nu \nu e^+ e^-} = \frac{1}{2M} \overline{|{\cal M}|^2} \int \dx\Phi_5(p_1;p_2,k_1,k_2,l_1,l_2).
    \label{eq:5bodydecay}
\end{equation}
Splitting the phase-space in two, we can instead calculate two subsequent three-body decays~\cite{ParticleDataGroup:2022pth} $N_2\rightarrow \bar{N_1} \nu \nu$ and $\nu \rightarrow e^-e^+\nu$ with
\begin{equation}
    \overline{|{\cal M}|^2} \dx\Phi_5 = \frac{1}{2\pi} \overline{|{\cal M}|^2} \dx\Phi_3(q;l_1,l_2,k_2) \dx\Phi_3(p_1;p_2,k_1,q) \dx q^2\,,
    \label{eq:phasespace_splitting}
\end{equation}
where the virtual neutrino has a ``mass'' of $q^2$. First, we perform a tensor decomposition 
\begin{equation}
    l_1^\mu l_2^\nu k_2^\rho = q^\mu g^{\nu\rho} t_1(q,l_1,l_2,k_2) + q^\mu q^\nu q^\rho t_2(q,l_1,l_2,k_2) + (\mu\leftrightarrow\nu\leftrightarrow\rho)\,,
\end{equation}
where $t_i(q,l_1,l_2,k_2)$ represent functions with all momenta contracted in scalar products. Thus, we can factorise the matrix element into one part depending on the first decay with momenta $p_1$, $p_2$, $k_1$ and one part depending on the second decay with $l_1$, $l_2$, $k_2$, where all momenta are contracted among themselves or with $q$. Eq.\,\eqref{eq:phasespace_splitting} can now be written as
\begin{align}
    \frac{1}{(2\pi) \Lambda^8} &\left(2(k_1\cdot p_1)(q\cdot p_2)+2(k_1\cdot p_2)(q\cdot p_1)-(p_1\cdot p_2)(q\cdot k_1)\right) \dx\Phi_3(p_1;p_2,k_1,q)\times \nonumber\\
    & 2 (k_2\cdot l_1)(l_2\cdot q) \dx\Phi_3(q;l_1,l_2,k_2) \times \frac{\dx q^2}{q^4} ,
\end{align}
and we can perform each three-body decay independently. For $\nu \rightarrow e^-e^+\nu$, we can assume that the mass of the decay products are negligible and arrive at
\begin{equation}
    \int 2 (k_2\cdot l_1)(l_2\cdot q) \dx\Phi_3(q;l_1,l_2,k_2) = \frac{q^6}{(2\pi)^3\cdot 48}\,.
\end{equation}
For $N_2\rightarrow \bar{N_1} \nu \nu$ we can expand in $\epsilon\ll 1$ and assume that $\bar{N_1}$ will be at rest at first order in $\epsilon$, while the neutrinos have a momentum of the order $\epsilon M$. With this simplification, the integral takes the form
\begin{align}
    \int \left(\right.2(k_1 & \cdot p_1)(q\cdot p_2)+2(k_1\cdot p_2)(q\cdot p_1)-(p_1\cdot p_2)(q\cdot k_1)\left.\right) \dx\Phi_3(p_1;p_2,k_1,q) = \nonumber \\
    = &\frac{M}{(2\pi)^3\cdot 40} \cdot \bigg[\sqrt{M^2\epsilon^2-q^2} \left(-9 M^2 \epsilon^2 q^2 + 2 M^4\epsilon^4 -8 q^4 \right) \nonumber \\
    & + 15 M \epsilon q^4 \ln{\frac{\sqrt{M^2\epsilon^2-q^2} + M\epsilon}{q}}\bigg]  \Theta(q^2) \Theta(\sqrt{q^2}-\epsilon M)\,. 
\end{align}
Now, the $q^2$ integration can be done with $q^2_\text{min}=0$ and $q^2_\text{max}=(\epsilon M)^2$ and we have
\begin{equation}
    \int \dx\Phi_5 \overline{|{\cal M}|^2} = \frac{ M (\epsilon M)^9}{30240\cdot (2\pi)^{7} \Lambda^8}\,.
\end{equation}
This result multiplied by $1/(2M)$ gives then the final decay width in Eq.\,\eqref{eq:5bodydecay_result}.

\subsection{Decay channel into photons}

In this section we give some details for the $N_2\rightarrow \bar{N}_1\nu  \nu\gamma\gamma$ decay discussed in Sec.~\ref{sec:decay_photons}. 
The decay involves a virtual Higgs boson decaying via $h^*\to\gamma\gamma$. This well-known process~\cite{ELLIS1976292,Gastmans:2011wh,Huang:2011yf,Marciano:2011gm,PhysRevD.104.053010} can be characterized by a contribution to the matrix element from Higgs decay given by~\cite{Marciano:2011gm}
\begin{align}
    {\cal M}_{h^*\rightarrow\gamma\gamma} = 
    \frac{2 e^2}{(4 \pi )^2 v_\text{EW}} F(q^2) \left(q_1 \cdot q_2 g^{\mu \nu }-q_1^{\nu } q_2^{\mu }\right) \epsilon _{\mu }(q_1) \epsilon _{\nu }(q_2)\,,
\end{align}
where $e$ is the electromagnetic gauge coupling constant, $q_i$ are the photon four-momenta, $q=q_1+q_2$ that of the Higgs, and $F(q^2)$ is a loop factor.
The virtuality $q^2$ takes the role of the Higgs ``mass'' for off-shell Higgs decay. Since the decay is dominated by $W$ and top-quark contributions, and $q^2= {\cal O}(\epsilon M)^2 \ll m_W^2, m_t^2$, the loop factor approaches a constant value corresponding to the heavy-top/$W$ limit. With these simplifications,  $|F|^2\simeq 27.3$.
The full squared matrix element for the process $N_2\rightarrow \bar N_1 \nu \nu h^* \rightarrow \bar N_1 \nu \nu \gamma \gamma$ is in the limit $q^2\ll m_h^2$ given by
\begin{align}
    \overline{|{\cal M}|^2}_{N_2\to\bar N_1\nu\nu\gamma\gamma} = \frac{\alpha_{\text{em}}^2 |F|^2 \left(q_1\cdot q_2\right)^2}{4\pi^2 m_h^4 v_{\text{EW}}^4} \overline{|{\cal M}|^2}_{N_2\to\bar N_1\nu\nu}\,,
\end{align}
where $\alpha_{\text{em}}=e^2/(4\pi)$ is the fine-structure constant, $m_h$ the Higgs mass, and the last factor stands for Eq.\,\eqref{eq:matrixelementsquared}. Similarly to before, the photon momenta $q_i$ can only be of the order $\mathcal{O}(\epsilon M)$ since most of the available energy in the $N_2$-decay is transferred to the rest mass of the daughter particle $\bar{N_1}$.  For $\epsilon\ll1$, the matrix element $\overline{|{\cal M}|^2}_{N_2\to\bar N_1\nu\nu\gamma\gamma}$ is suppressed by a factor $\epsilon^4$ compared to  $\overline{|{\cal M}|^2}_{N_2\to\bar N_1\nu\nu}$. Furthermore, the five-body phase-space leads to an additional suppression by a factor $\epsilon^4$ as compared to the three-body decay, as well as an additional $1/\pi^2$ factor. Overall, this allows us to estimate the branching fraction as given in Eq.\,\eqref{eq:branchingratioHiggs}, where we counted $|F|^2/\pi^2$ as order unity for simplicity.

\section{Dark matter production via freeze-in}
\label{app:freeze-in}
For the DM production rate we have to solve
\begin{equation}
\gamma_{N_1N_2}\equiv 2 \int \dx\Pi_{k_1} \dx\Pi_{k_2} \dx\Pi_{p_1} \dx\Pi_{p_2} (2\pi)^4 \delta^{(4)}(k_1+k_2-p_1-p_2) \overline{|{\cal M}|^2}_{\nu\nu\rightarrow N_1 N_2} f_\nu(k_1) f_\nu(k_2)\,,
\end{equation}
where $f_\nu(k)=1/(e^{k/T}+1)$ since neutrinos belong to the SM thermal bath at the relevant temperatures.  Following~\cite{Edsjo:1997bg} the integrals over $N_{1,2}$ momenta $p_{1,2}$ and neutrino momenta $k_{1,2}$ can be evaluated, leaving an integration over the squared center-of-mass energy $s=(k_1+k_2)^2=(p_1+p_2)^2$ as well as the angle between \emph{e.g.}~the spatial momenta of $k_1$ and $p_1$,
\begin{equation}\label{eq:gammaN1N2Edsjo}
\gamma_{N_1N_2} = \frac{T}{8\cdot 32\pi^6} \int \dx s \, \frac{p_{\nu\nu} p_{N_1 N_2}}{\sqrt{s}} \, K_1\left(\frac{\sqrt{s}}{T}\right) \, \int\dx\Omega\, \overline{|{\cal M}|^2}_{\nu\nu\rightarrow N_1 N_2}\,,
\end{equation}
with $K_1$ being the modified Bessel function of the second kind of order one and
\begin{equation}
    p_{ij} \equiv \frac{\sqrt{s-(m_i+m_j)^2}\sqrt{s-(m_i-m_j)^2}}{2\sqrt{s}}\,.
\end{equation}
Using  Eq.\,\eqref{eq:nunuNN} we obtain
\begin{equation}
    \overline{|{\cal M}|^2}_{\nu\nu\rightarrow N_1 N_2} = \frac{\vew^4}{4\Lambda^8} \left( 2 (k_1\cdot p_2) (k_2\cdot p_1) + 2 (k_1\cdot p_1) (k_2\cdot p_2) - (k_1\cdot k_2) (p_1\cdot p_2)\right)\,.
\end{equation}
We assume $m\simeq M$, which is a valid approximation for small mass splitting $M-m\simeq\epsilon M\ll M$ since only temperatures $T\gtrsim M\gg\epsilon M$ are relevant for freeze-in. Then, we find
\begin{equation}
    \int\dx\Omega\, \overline{|{\cal M}|^2}_{\nu\nu\rightarrow N_1 N_2} = \frac{\pi}{12}\frac{\vew^4  }{ \Lambda^8}s (2M^2+s)\,.
\end{equation}
Furthermore, $p_{\nu\nu} = {\sqrt{s}}/{2}$ and  $p_{N_1 N_2} = {\sqrt{s-4M^2}}/{2}$ as well as $s\geq 4M^2$ in the limit of small mass splitting $m\simeq M$. Performing the remaining integral over $s$ from $4M^2$ to $\infty$ in Eq.\,\eqref{eq:gammaN1N2Edsjo} then reproduces Eq.\,\eqref{eq:gammaN1N2}.

Freeze-in via $\nu\nu\to N_1N_2$ leads to equal initial abundances of $N_1$ and $N_2$.
One may wonder whether conversion processes may alter the relative abundances. Clearly, the decay process  $N_2\rightarrow \bar N_1\nu\nu$ itself is irrelevant in the Early Universe due to the cosmological DCDM lifetime we consider. However, conversions could also be mediated by scatterings of \emph{e.g.}~the form $N_2 \bar\nu \to\bar N_1 \nu$. Nevertheless, at temperatures $T\gg \epsilon M$, the rate of this and its inverse process is practically identical due to the mass splitting being negligible at these scales. Therefore, they would not change the relative abundance of $N_2+\bar N_2$ versus $N_1+\bar N_1$ particles even if they would occur at sizeable rates. For $T\ll \epsilon M$, the mass splitting becomes relevant, and would lead to a preference of the de-excitation process $N_2 \bar\nu \to\bar N_1 \nu$ over its inverse. However, we checked that its rate is strongly suppressed compared to the Hubble rate at these low temperatures. In conclusion, we can neglect conversion processes.

Lastly, the equal initial abundances of $N_1+\bar N_1$ and $N_2+\bar N_2$ also causes only $50\%$ of DM to decay which we have to account for in our previous analyses.
From the point of view of cosmological observations, the initial $N_1+\bar N_1$ population acts as a component of stable cold dark matter (SCDM), since its velocity distribution inherited by the freeze-in process is negligibly small at around the recombination epoch. Therefore, we need to consider in general three distinct DM populations, being DCDM, WDM and SCDM, corresponding to $N_2+\bar N_2$, $N_1+\bar N_1$ produced via DCDM decay, and the initial $N_1+\bar N_1$ population, respectively. At any given time the fraction of DM in the form of DCDM, WDM and SCDM is $0.5e^{-t/\tau}$, $0.5(1-e^{-t/\tau})$ and $0.5$, respectively (assuming $m\simeq M$,\emph{i.e.}~$\epsilon\ll1$, as before).
Since the DCDM lifetimes $\tau\simeq 10^2$\,Gyrs that we are interested in are somewhat above the age of the Universe $t_0$. It is not necessary to track the three components separately in this case, but it is sufficient to consider the total CDM density given by the sum of DCDM and SCDM populations. In particular,
\begin{align}
    & \rho_{\text{cdm}} = \rho_{N_1+\bar N_1,\text{initial}} + \rho_{N_2+\bar N_2} = \frac{1}{2}\rho_0 a^{-3} + \frac{1}{2}\rho_0 a^{-3} e^{-t_0/\tau} \approx \rho_0 a^{-3} \left( 1-\frac{t_0}{2\tau} \right)\,, \nonumber\\
    & \rho\wdm = \rho_{N_1+\bar N_1,\text{decay}} = \frac{1}{2}\rho_0 a^{-3} \left( 1-e^{-t_0/\tau} \right) \approx \rho_0 a^{-3} \frac{t_0}{2\tau}\,,
\end{align}
where we expanded for $\tau\gg t_0$ in the last expressions in each line, and $\rho_0$ stands for the total DM density today. This can be compared to a corresponding model  in which initially only the heavier state is present, and with lifetime denoted by $\tau'$. The populations of cold and warm dark matter are then given by
\begin{align}
    & \rho_{\text{cdm}} = \rho_0 a^{-3} e^{-t_0/\tau'} \approx \rho_0 a^{-3} \left( 1-\frac{t_0}{\tau'} \right)\,, \nonumber\\
    & \rho\wdm = \rho_0 a^{-3} \left( 1-e^{-t_0/\tau'} \right) \approx \rho_0 a^{-3} \frac{t_0}{\tau'}\,.
\end{align}
Thus we see that for $\tau\gg t_0$ both setups can be mapped to each other when identifying $\tau'\equiv 2\tau$.

\section{Dark matter-neutrino scatterings}
\label{app:NuDM}

\begin{figure}[t]
    \centering
    \includegraphics[width=0.6\textwidth]{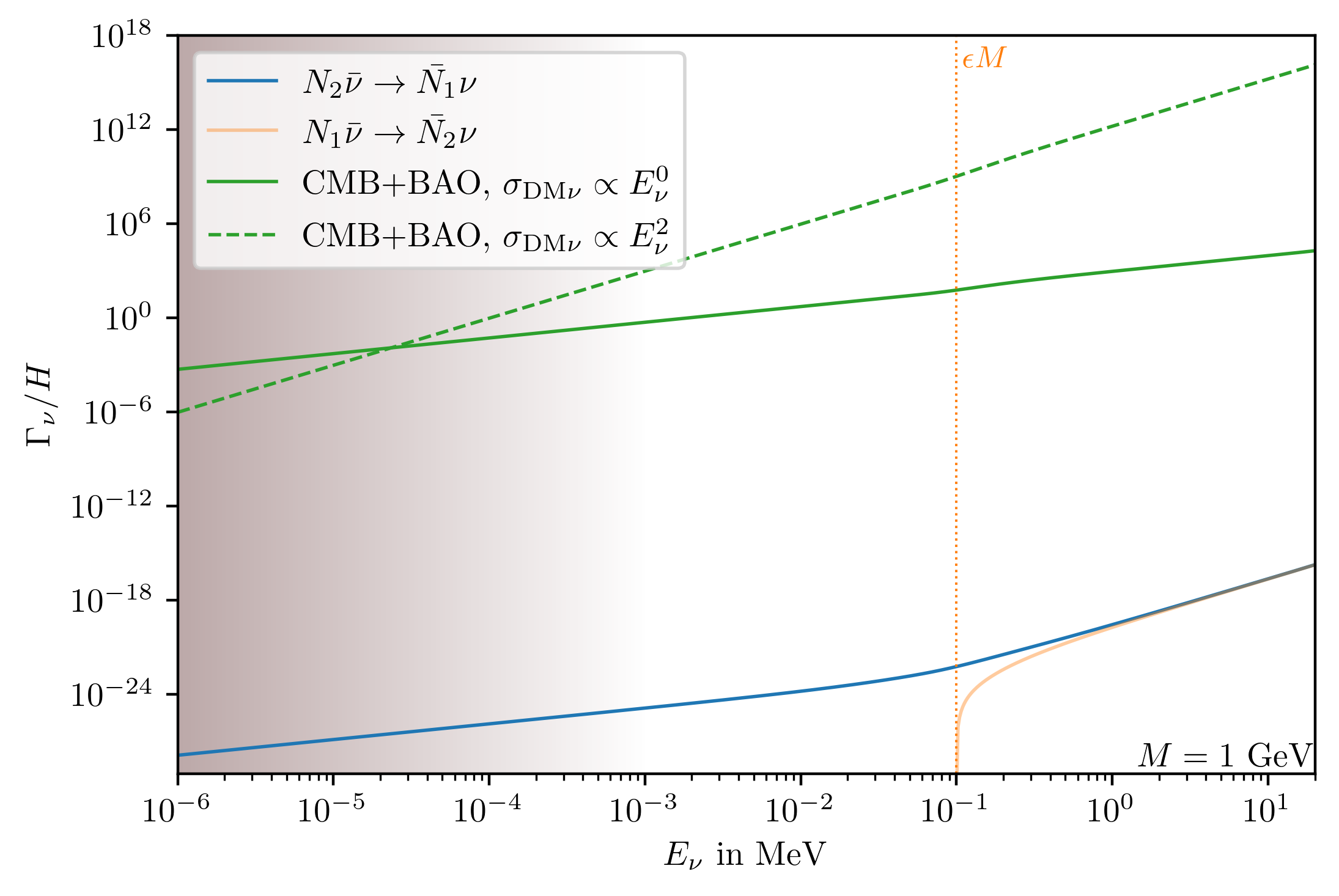}
    \caption{Ratio of the neutrino scattering rate and Hubble parameter versus the typical thermal neutrino energy $E_\nu\sim T$. In blue, the scattering rate for $N_2\bar\nu\rightarrow\bar{N_1}\nu$ 
    is shown for typical parameter values $M=1$\,GeV, $\epsilon=10^{-4}$ and $\tau = 100$\,Gyrs.
    For comparison, we display limits on the elastic neutrino-DM cross section $\sigma_{\text{DM}\nu}$ from~\cite{Brax:2023tvn} based on CMB and BAO data. The shaded region on the left indicates the energy regime that is relevant for CMB constraints.}
    \label{fig:DMlowenergyscattering}
\end{figure}

Calculating the cross section for the $N_2\bar \nu\to N_1\nu$ and analogously for the inverse process yields
\begin{align}\label{eq:DMneutrinoscattering}
    \sigma_{N_2 \bar\nu \rightarrow \bar{N_1}\nu} &= \frac{v_{EW}^4}{256 \pi \Lambda^8} \frac{(s-m^2)^2}{s} \quad \text{and}\\
    \sigma_{N_1 \bar\nu \rightarrow \bar{N_2}\nu} &= \frac{v_{EW}^4}{256 \pi \Lambda^8} \frac{(s-M^2)^2}{s}\,,
\end{align}
where $s$ is the square of the center-of-mass energy and we assumed massless neutrinos.
As long as the effective four-fermion interaction is applicable (\emph{i.e.}~for $\sqrt{s}\ll\Lambda$), we thus find a scaling with $s$, as expected on dimensional grounds. For $N_2\bar \nu\to N_1\nu$ one has $s\simeq M^2 + 2 M E_\nu$ when assuming the $\bar N_2$ to be non-relativistic. This can be assumed in the Early Universe for temperatures $T\ll M$. In this regime, the cross section depends on the masses and the neutrino energy, and thus temperature, with $E_\nu \sim T$. For $E_\nu\gg \epsilon M$, $\sigma_{N_2 \bar\nu \leftrightarrow \bar{N_1}\nu} \propto s \propto E_\nu^1$, and the mass splitting between $N_2$ and $N_1$ can be neglected, such that the scattering can be considered effectively as elastic. For $E_\nu \ll \epsilon M$, we have $\sigma_{N_2 \bar\nu \rightarrow \bar{N_1}\nu} \propto (\epsilon M)^2$ for the de-excitation, while the backward reaction is kinematically forbidden. In this regime, the scattering is strongly inelastic since the mass splitting of the DM particles dominates over the neutrino energy.

Assuming that DM is dominantly comprised of the heavier $N_2$ particles in the Early Universe, the neutrino scattering rate for $T\ll M$ can be estimated as
\begin{equation}
    \Gamma_{\nu} = \sigma_{N_2 \bar\nu \rightarrow \bar{N_1}\nu} \cdot v \cdot n_{N_2}\,,
\end{equation}
where $n_{N_2} = \rho_{\text{DM}}/M$ with $\rho_{\text{DM}}\propto a^{-3}\propto T^3$ being the non-relativistic DM density and relative velocity $v\approx 1$. This can be compared to the usual Hubble rate $H\propto T^2$ to quantify whether the scattering may impact neutrino free-streaming. We show the ratio $\Gamma_\nu/H$ versus the neutrino energy $E_\nu\sim T$ in Fig.\,\ref{fig:DMlowenergyscattering} for $M=1$\,GeV, $\epsilon=10^{-4}$ and $\tau=100$\,Gyrs (blue line). For $E_\nu\ll\epsilon M$ one has $\sigma_{N_2 \bar\nu \rightarrow \bar{N_1}\nu}  \propto E_\nu^0$ such that $\Gamma_\nu/H\propto T\sim E_\nu$ while for $E_\nu\gg\epsilon M$ the scaling $\sigma_{N_2 \bar\nu \rightarrow \bar{N_1}\nu}  \propto E_\nu^1$ implies $\Gamma_\nu/H\propto T^2\sim E_\nu^2$. For illustration, we also show the scattering rate that would result from the inverse process $N_1 \bar\nu \rightarrow \bar{N_2}\nu$ when assuming all of DM would be in the form of $N_1$ (faint orange line). While we do not consider this scenario further, we note that both rates become equal for $E_\nu\gg\epsilon M$, illustrating the effective elasticity in this limit.
We also display constraints on neutrino-DM scattering from CMB and BAO observations~\cite{Brax:2023tvn}  in Fig.\,\ref{fig:DMlowenergyscattering}. We stress that these constraints are obtained assuming purely \emph{elastic} scattering, and are thus comparable to the model studied here only for $E_\nu\gg \epsilon M$. In this context, constraints are commonly expressed in terms of the dimensionless ratio
\begin{equation}
    u_{\nu\text{DM}} = \frac{\sigma_{\nu\text{DM}}}{\sigma_T} \left( \frac{M}{100 \text{GeV}} \right) ^{-1}\,,
\end{equation}
where $\sigma_T$ is the Thompson cross section. In~\cite{Brax:2023tvn} a dependence $\sigma_{\nu\text{DM}}\propto T^k$ was assumed, finding upper limits $\log_{10}(u_{\nu\text{DM}})\lesssim -5$ for $k=0$ and $\log_{10}(u_{\nu\text{DM}})\lesssim -15$ for $k=2$. We show these two upper limits in Fig.~\ref{fig:DMlowenergyscattering} in green solid and dashed lines, respectively. They both fall below $\Gamma_\nu/H \lesssim 1$ at energy scales $\mathcal{O}(0.1)$\,keV. This can be related to the fact that the small angular scales used in the CMB analysis of~\cite{Brax:2023tvn} enter the horizon when the temperature of the thermal bath was around these energy scales. We observe that the neutrino-DM scattering rate predicted by the DCDM model under consideration is suppressed by many orders of magnitude in this temperature regime and consequently can be neglected.

\section{Decaying light dark matter}
\label{app:lowmass}

So far we focused on the DM mass range $M\simeq\mathcal{O}(\mathrm{GeV})$, for which the suppression scale $\Lambda$ of the effective operator mediating DCDM decay lies in the multi-TeV range. Here, we discuss the low-mass regime with $M\simeq\mathcal{O(\text{MeV})}$. Using Eq.\,\eqref{eq:Lambda} to rewrite $\Lambda$ as
\begin{equation}
    \Lambda \simeq 160 \mathrm{GeV} \left( \frac{\tau}{100 \mathrm{Gyrs}} \left(\frac{\epsilon M}{10^{-3}\cdot 1\mathrm{MeV}}\right)^5 \right)^{1/8}\,, 
\end{equation}
implies that the typical interaction strength $\propto 1/\Lambda^8$ is enhanced. For even lower masses one would have $\Lambda\ll v_{\text{EW}}$, which indicates that an EFT approach within the broken phase should be used. Thus, while the low-mass regime can boost some signatures, it is constrained by the validity of the EFT description.

\begin{figure}[t]
    \centering
    \includegraphics[width=0.6\textwidth]{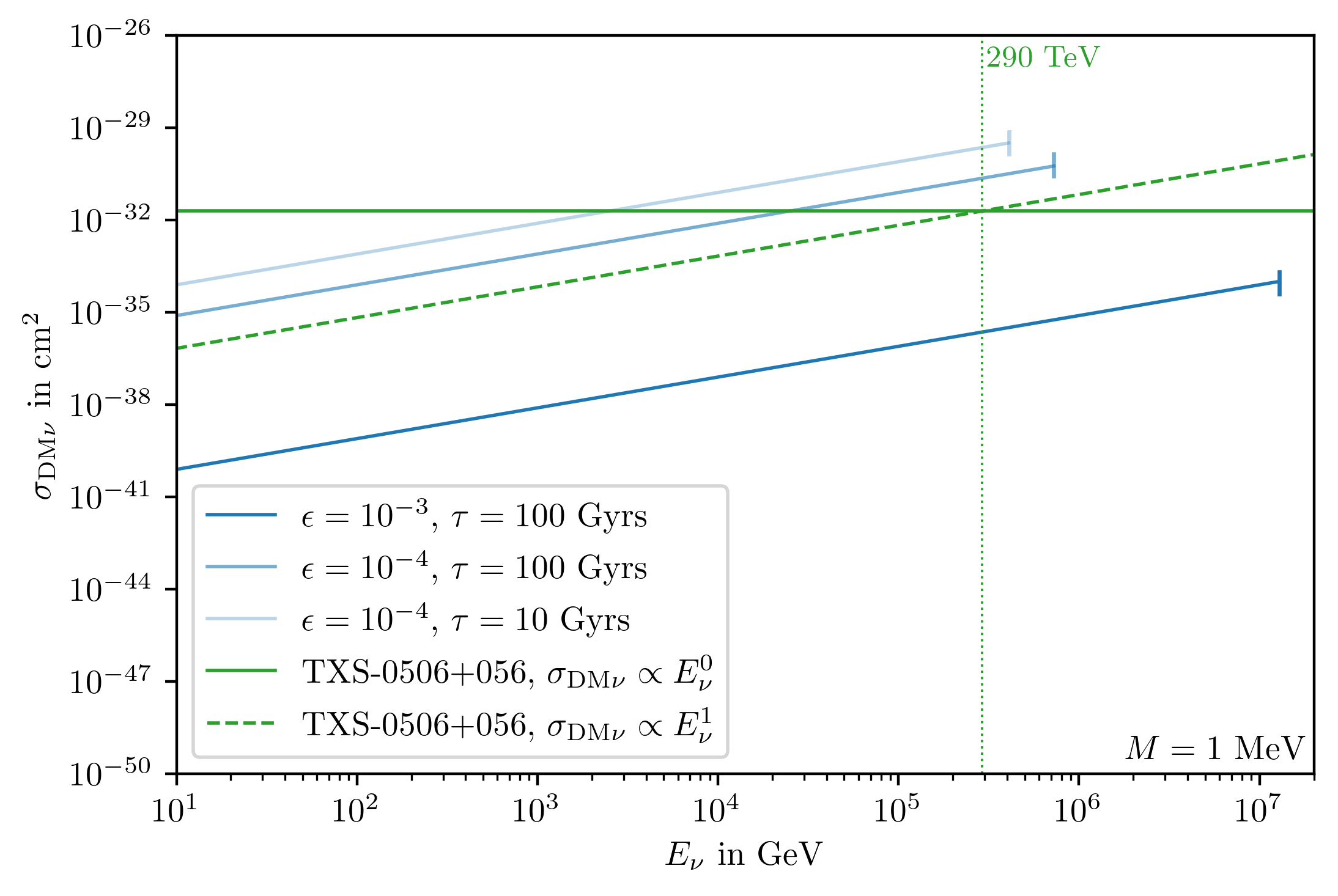}
    \caption{Cross section of neutrino-DM scattering  for $M=1$\,MeV and different $\epsilon$ and $\tau$ values in shades of blue. In green, constraints derived from IceCube observations of high-energy neutrinos observed from the direction of TXS-0506+056 are shown~\cite{Ferrer:2022kei}.}
    \label{fig:highenergyneutrinosMeV}
\end{figure}

\begin{figure}[t]
    \centering
    \includegraphics[width=0.6\textwidth]{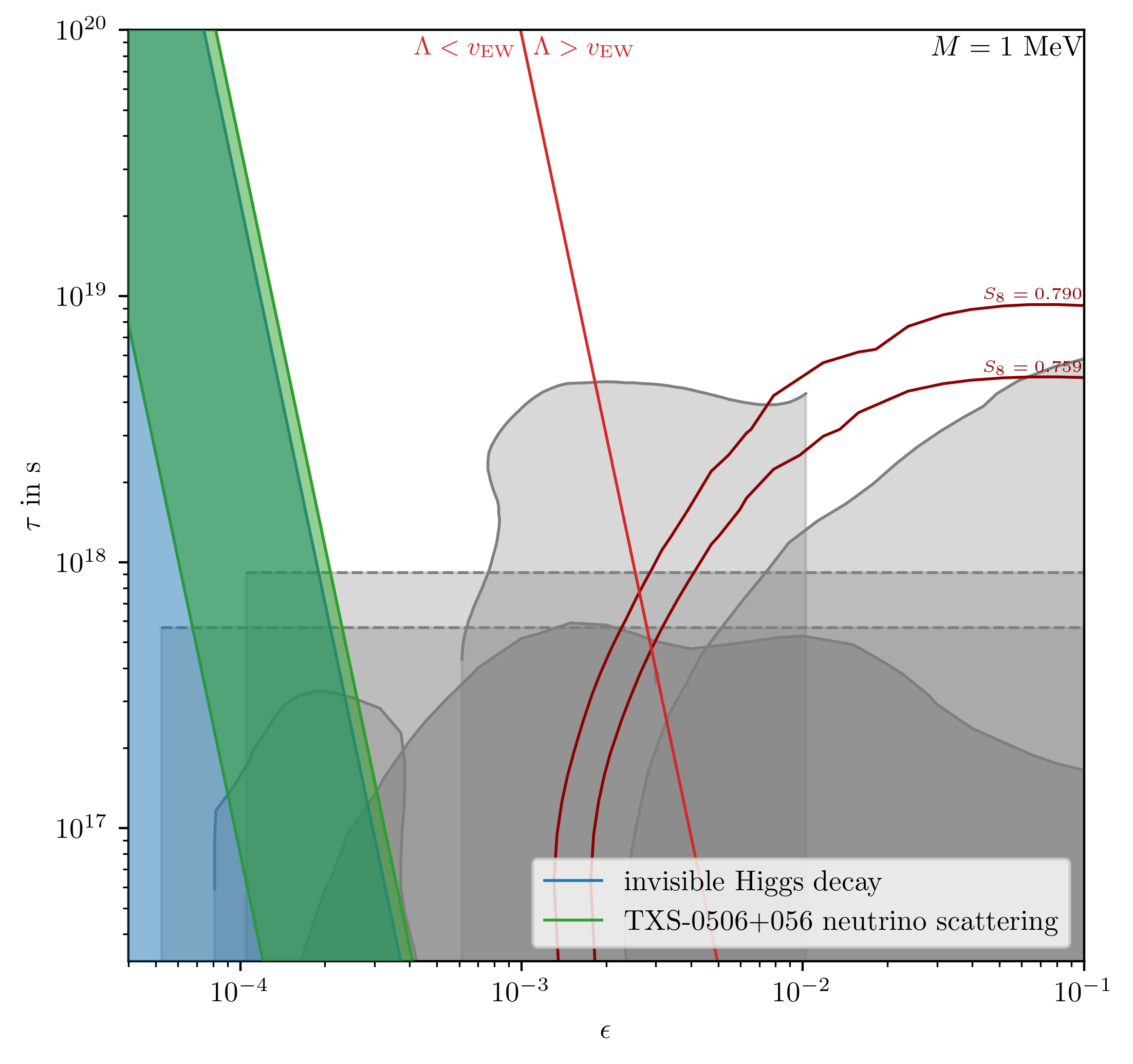}
    \caption{Exclusion plot for $M=1$\,MeV showing the cosmological constraints in gray, the constraint from the invisible Higgs decay in blue and the IceCube constraint from high-energy neutrino scattering in green. The red line shows when $\Lambda$ becomes smaller than $v_{\text{EW}}$.}
    \label{fig:exclusionlowmasses}
\end{figure}

In Fig.~\ref{fig:highenergyneutrinosMeV} we show the neutrino-DM scattering cross section for $M=1$\,MeV and
see that IceCube limits can put relevant constraints on DCDM in the low-mass regime. The cross section is shown only for $\sqrt{s}<\Lambda$ to ensure EFT validity and additionally, we checked that the cross section satisfies the unitarity bound. The resulting constraints can be seen in green in Fig.~\ref{fig:exclusionlowmasses}.
Similarly, for invisible Higgs decay, we find non-trivial constraints when assuming that the EFT description remains valid for $\Lambda<v_{\text{EW}}$, as illustrated in blue in Fig.~\ref{fig:exclusionlowmasses}. 
However, in red we also show the line where $\Lambda=v_{\text{EW}}$. Since the momentum transfer for the invisible Higgs decay is given by the scale of the Higgs mass which is comparable to $v_{\text{EW}}$, the bound is clearly already in the regime where the EFT is non-sufficient. Thus, a UV completion would be required to assess constraints from invisible Higgs decay within the low-mass regime.

\begin{figure}[t]
    \centering
    \includegraphics[width=0.6\textwidth]{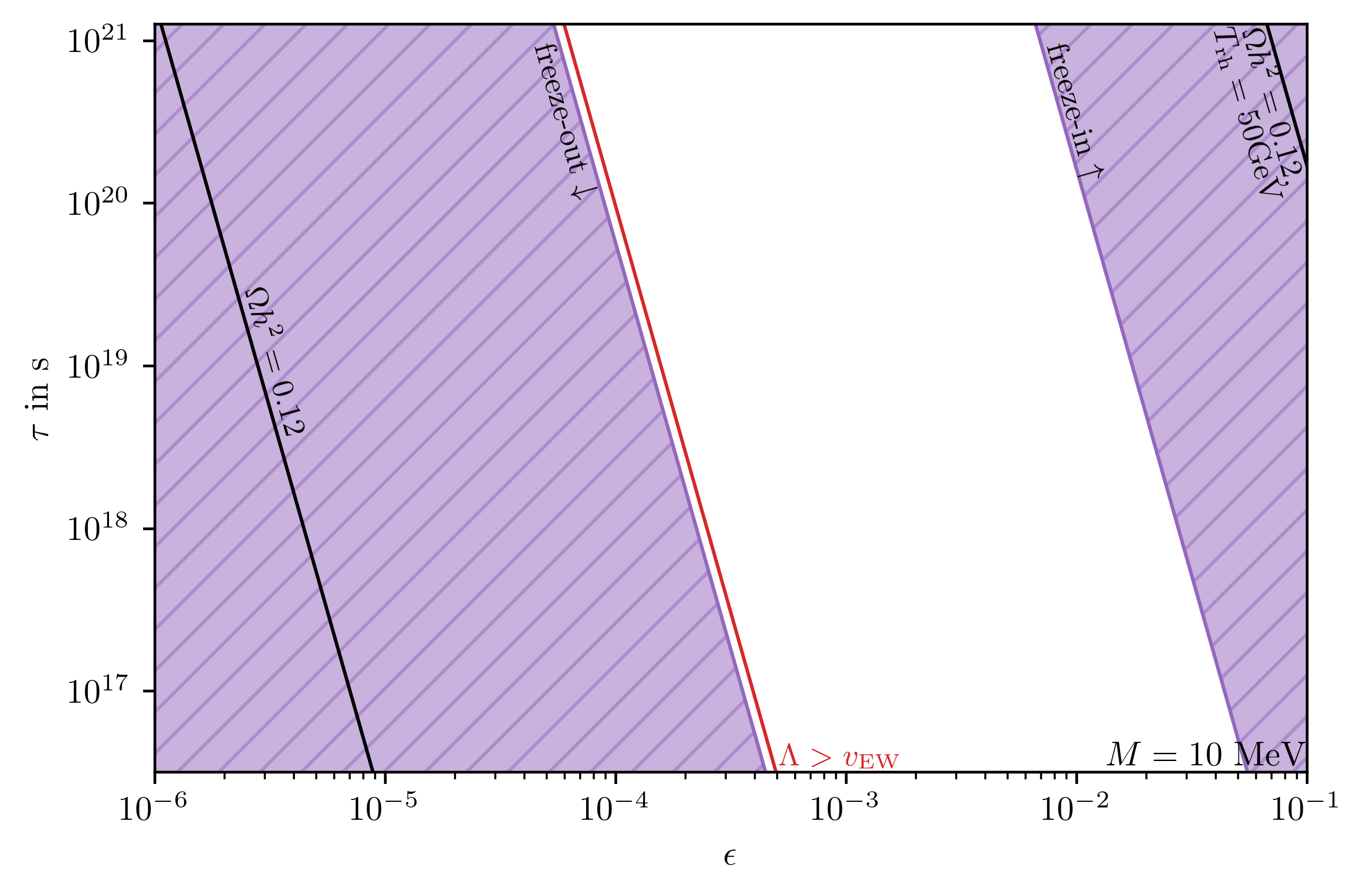}
    \caption{Regimes where production can occur via freeze-out (on the left side) as well as freeze-in (on the right) for an exemplary reheating temperature of $T_\text{rh}=50$\,GeV. In both cases, the black line indicates where a relic abundance of $\Omega h^2 = 0.12$ can be reached.}
    \label{fig:freezeout_10MeV}
\end{figure}

Lastly, the production mechanism explained in Sec.~\ref{sec:Production} is also affected by the lowered mass because freeze-in requires a sufficiently weak  coupling. Otherwise, the DM particles thermalize, and freeze-out occurs instead which happens for relevant values of $\epsilon$ and $\tau$ for lower masses of $M\simeq10$\,MeV. To quantify this regime, we require that the production rate at a temperature equal to the DM mass has to be larger than the Hubble parameter $\left. n_\nu \langle \sigma v\rangle (T=M) \right\vert_{\text{freeze-out}}  > H(T=M)$. For the freeze-in regime, we conservatively require that the production rate at the reheating temperature is smaller than the Hubble rate $\left. n_\nu \langle \sigma v\rangle (T=T_{\text{rh}}) \right\vert_{\text{freeze-in}}  < H(T=T_{\text{rh}})$, so that the back-reaction can always be neglected.
In Fig.~\ref{fig:freezeout_10MeV}, both the freeze-in regime for an exemplary choice $T_\text{rh}=50$\,GeV and the freeze-out regime are shown, for $M=10$\,MeV.
Solving the Boltzmann equation for freeze-out to determine the relic abundance of DM, results in the black line on the left, while the one on the right accounts for the freeze-in solution for the selected reheating temperature. We note that the freeze-out region corresponds to $\Lambda<v_{\text{EW}}$, while all energy scales relevant for freeze-out are well below $\Lambda$.
Nevertheless, a complete study of the phenomenology within the low-mass regime requires a UV completion, which is left to future work.

\providecommand{\href}[2]{#2}\begingroup\raggedright\endgroup

\end{document}